%% file: paper.tex
\documentclass[sigconf]{acmart}

\usepackage{epstopdf}
\usepackage{algpseudocode}
\usepackage{amsfonts}
\usepackage{algorithm}
\usepackage{mathtools}
\usepackage{tikz}

\usepackage{booktabs}
\usepackage{tabularx}
\usepackage [english]{babel}
\algrenewcommand{\algorithmiccomment}[1]{\textcolor{ACMDarkBlue}{// #1}} 
\newcommand*\Let[2]{\State #1 $\gets$ #2}
\newcommand*\amp{\text{ \& }}

\renewcommand\footnotetextcopyrightpermission[1]{} 
\newcommand\blfootnote[1]{%
  \begingroup
  \renewcommand\thefootnote{}\footnote{#1}%
  \addtocounter{footnote}{-1}%
  \endgroup
}

\makeatletter
\let\OldStatex\Statex
\renewcommand{\Statex}[1][3]{%
  \setlength\@tempdima{\algorithmicindent}%
  \OldStatex\hskip\dimexpr#1\@tempdima\relax}
\makeatother

\ifpdf%
  \DeclareGraphicsExtensions{.eps,.pdf,.png,.jpg}
\else
  \DeclareGraphicsExtensions{.eps}
\fi

\acmConference[PASC'18]{Platform for Advanced Scientific Computing}{July 02--04}{Basel, Switzerland}
\acmISBN{}
\acmPrice{}

\begin{document}

\title{A Parallel Solver for Graph Laplacians}

\author{Tristan Konolige}
\affiliation{University of Colorado Boulder}
\email{tristan.konolige@colorado.edu}
\author{Jed Brown}
\affiliation{University of Colorado Boulder}

\begin{abstract}
  Problems from graph drawing, spectral clustering, network flow and graph partitioning can all be expressed in terms of graph Laplacian matrices.
  There are a variety of practical approaches to solving these problems in serial.
  However, as problem sizes increase and single core speeds stagnate, parallelism is essential to solve such problems quickly.
  We present an unsmoothed aggregation multigrid method for solving graph Laplacians in a distributed memory setting.
  We introduce new parallel aggregation and low degree elimination algorithms targeted specifically at irregular degree graphs.
  These algorithms are expressed in terms of sparse matrix-vector products using generalized sum and product operations.
  This formulation is amenable to linear algebra using arbitrary distributions and allows us to operate on a 2D sparse matrix distribution, which is necessary for parallel scalability.
  Our solver outperforms the natural parallel extension of the current state of the art in an algorithmic comparison.
  We demonstrate scalability to 576 processes and graphs with up to 1.7 billion edges.
\end{abstract}


\maketitle

\blfootnote{Our solver is available at \texttt{https://github.com/ligmg/ligmg}}

\section{Introduction}\label{sec:intro}

Graph Laplacians arising from large irregular degree graphs present a variety of challenges for typical multigrid solvers.
Irregular graphs often present a broad degree distribution and relatively low diameter.
Although coarse grid complexity is relatively predictable and can be controlled directly in multigrid methods for PDE problems (typically bounded degree and large diameter), that is not the case for irregular graphs.
Consequently, methods such as smoothed aggregation lead to loss of sparsity in coarse grids~\citep{LAMG}.
Standard coarsening strategies for parallel aggregation either use some form of vertex partitioning or a maximal independent set~\citep{mlguide,bell2012exposing,brannick2013parallel}.
Vertices in irregular graphs cannot even be partitioned among processes (we use an edge distribution for load balancing reasons), much less partitioned into aggregates.
A maximal independent set is not viable for irregular graphs or those with small diameter.
For example, a single vertex with degree of one million will produce unacceptably rapid coarsening.
Even if high-degree vertices are removed, social network graphs such as the Facebook graph have very small maximal independent sets~\citep{backstrom2012four,boldi2012four}, which also lead to unacceptably rapid coarsening.
To avoid this, any viable coarsening scheme must permit adjacent root vertices (as MIS forms the smallest possible aggregates without adjacent roots).
We present a new parallel algorithm for choosing aggregates that accurately captures low energy components while maintaining coarse grid sparsity.
We also develop a parallel algorithm for finding and eliminating low degree vertices, a technique introduced for a sequential multigrid algorithm by Livne and Brandt~\citep{LAMG}, that is important for irregular graphs.

While matrices can be distributed using a vertex partition (a 1D/row distribution) for PDE problems, this leads to unacceptable load imbalance for irregular graphs.
We represent matrices using a 2D distribution (a partition of the edges) that maintains load balance for parallel scaling \citep{CombBLAS}.
Many algorithms that are practical for 1D distributions are not feasible for more general distributions.
Our new parallel algorithms are distribution agnostic.

\section{Background}\label{sec:background}

\subsection{Graph Laplacian Matrix}

Graphs arise in many areas to describe relationships between objects. They can
be people and friendships in a social network, computers and network
connections in a local area network, cities and the roads that connect them,
etc. We start with a matrix representation of a graph.

A weighted graph $G = (V,E,w)$ (where $V$ are vertices, $E$ are edges and $w$
are edge weights) can be expressed as an adjacency matrix $A$:
$$
A_{ij} = \begin{cases}
  w_{ij} & \text{if } (i,j) \in E\\
  0 & \text{otherwise}
\end{cases}
$$

The Laplacian matrix $L$ can then be expressed as:
\begin{align*}
  L &= D - A \\
  D_{ij} &= \begin{cases}
    \sum_u A_{uj} & \text{if } i = j \\
    0 & \text{otherwise}
  \end{cases}
\end{align*}

In this paper we consider positively weighted ($w \ge 0$), undirected ($(i,j)
\in E \iff (j,i) \in E$ and $w_{ij} = w_{ji}$) graphs. Graph Laplacians of this
type have a couple of properties:

\begin{itemize}
    \item Column and row sums are zero.
    \item Off diagonal entries are negative.
    \item Diagonal entries are positive.
    \item $L$ is symmetric positive semi-definite.
\end{itemize}

In this paper, we assume that the graph $G$ is connected.
A connected graph is one where each vertex can reach every other vertex by traversing edges.
For example, in a connected social network, every person has a series of friendships that link them to any other person.
Assuming a connected graph simplifies our setup phase because we do not need to keep track of each component.

For a connected graph, the corresponding Laplacian matrix has a null space spanned by the constant vector.
The dimension one nullspace is easier to keep track of and orthogonalize against.
Given the eigensystem $ L u_i = \lambda_i u_i $, the eigenvalues  $0 = \lambda_0 \leq \lambda_1 \leq \lambda_2 \leq ... \leq \lambda_{n-1}$ are non-negative and real.
The number of eigenvalues $\lambda_i=0$ is equal to the number of connected components in the graph.
Because we are only considering connected graphs, $\lambda_1 > 0$.
The eigenvector, $u_1$, associated with the second smallest eigenvalue approximates the sparsest cut in the graph.
The sign of $u_1$ determines which side of the cut each vertex belongs to \citep{spielman2007spectral}.

Some applications of the graph Laplacian include:
\begin{itemize}
  \item \textbf{Graph partitioning} As mentioned above, the second smallest eigenvalue of the graph
    Laplacian approximates the sparsest cut in the graph.
  \item \textbf{Graph drawing} Like graph partitioning, spectral graph
    drawing relies on finding eigenvalues and eigenvectors of the Laplacian to
    embed the graph in a two dimensional space for visualization. Each vertex, $i$, is placed at $(u_1(i), u_2(i))$. Other graph drawing techniques, such as the maxent-stress model, rely on solving the graph Laplacian \citep{maxent-stress}.
  \item \textbf{PDEs on unstructured meshes} Some discretizations of partial differential equations on unstructured meshes result in Laplacian matrices \citep{spielman_laplacian}.
    If the mesh has large variation in vertex degree, solution techniques for arbitrary graph Laplacians might outperform those designed for fixed degree or more structured meshes.
  \item \textbf{Electrical flow} A circuit of resistors with current inputs and sinks between them can be modeled as a graph Laplacian.
    Each vertex in the graph is a current input or sink, and edges correspond to resistors with resistance equal to reciprocal resistance.
    If $r$ is a vector where $r_i$ is the current input or draw, then solving $p = L^{-1} r$ gives the potential $p_i$ at vertex $i$ \citep{spielman_laplacian}.
\end{itemize}

See Spielman's article on applications of the graph Laplacian for more details \citep{spielman_laplacian}.

All these applications (either directly or indirectly) require applying the action of the inverse of the graph Laplacian ($L^{-1}$). We refer to computing $x$ in $L x = b$ as solving the graph Laplacian $L$. We refer to a method of solving $L x = b$ as a solver. To compute small eigenvectors and eigenvalues for applications such as graph partitioning, an eigensolver must be used. Typically, these eigensolvers iterate on the inverse matrix (in this case $L^{-1}$) or on a shifted inverted matrix. The majority of time is often spent in the linear solve (i.e., computing $L^{-1}x$).

There are many different kinds of graphs: road networks, computer networks, social networks, power grids, document relations, etc.
We focus our performance evaluation on social network graphs because they often provide the most difficulty for existing multigrid solvers.
Structured graphs, such as PDEs discretized on a grid, provide more opportunities for a solver to cut corners.
Social networks have a couple of properties that make them more difficult to solve than other graphs:

\begin{itemize}
  \item Social network graphs are sparse: the number of edges is roughly a
    constant factor of the number of vertices.
  \item Edges in these networks follow a power-law degree distribution (these
    graphs are called scale-free). A small number of vertices have a large
    number of neighbors, whereas the rest have a
    relatively small degree. This presents a challenge when determining how to
    distribute work in parallel.
  \item Social network graphs have high connectivity. Most vertices can be
    reached in a small number of hops from any other vertex. This causes large
    fill in for techniques such as LU factorization and smoothed aggregation
    multigrid.
\end{itemize}

\subsection{Related Work}\label{sec:relwork}

A variety of different solvers that have been proposed for solving
large sparse SPD systems. Some are general purpose solvers, whereas others are
tailored specifically for graph Laplacians.

\subsubsection{Direct Solvers}

A variety of direct solvers capable of solving sparse systems. Some of these
solvers, such as SuperLU\_DIST~\citep{superlu} and MUMPS~\citep{MUMPS01}, function in distributed memory.
However, none of these solvers performs well on large graph Laplacian systems because ``small'' vertex separators~\citep{george81} do not exist. Because
we are interested in very large systems, we require that our solver scales
linearly in the number of nonzeros in the matrix, which these direct solvers do
not.

\subsubsection{Simpler Preconditioners}

Iterative solvers such as conjugate gradients, coupled with simple preconditioners like Jacobi or incomplete Cholesky, sometimes perform well on highly irregular graphs.
For example, Jacobi (diagonal) preconditioning is often sufficient for social network graphs with small diameter and Incomplete Cholesky may provide more robustness, but tends to exhibit poor parallel scalability.

\subsubsection{Theoretical Solvers}

A variety of theoretical Laplacian solvers have been proposed in literature starting with Spielman and Teng's 2003 paper \citep{ST03}.
To our knowledge, no working implementation of this algorithm exists.
Since then, many more theoretical linear solvers have been proposed.
Most use either a support graph or low stretch spanning tree sparsifier as a preconditioner for a Krylov iterative solver.
Several serial implementations of these ideas exist in ``Laplacians.jl'' a package written by Daniel Spielman \citep{laplaciansjl}.

Kelner et al. later proposed a simple and novel technique with a complexity
bound of $O(m\log^2{n}\log{\log{n}}\log{(\epsilon^{-1})})$~\citep{Kel13}($n$ is the number of vertices, $m$ is the number of edges, $\epsilon$ is the solution tolerance). An
implementation of this algorithm exists but appears to not be practical (as of yet) \citep{BomanDG15}.

\subsubsection{Practical Serial Solvers}

Three practical graph Laplacian solvers have been proposed: Koutis and Miller's
Combinatorial Multigrid (CMG) \citep{CMG}, Livne and Brandt's Lean Algebraic
Multigrid (LAMG) \citep{LAMG}, and Napov and Notay's Degree-aware Rooted
Aggregation (DRA) \citep{DRA}. All use multigrid techniques to solve the
Laplacian problem.

Combinatorial Multigrid, like much of the theoretical literature, takes a graph
theoretic approach.  It constructs a multilevel preconditioner using clustering
on a modified spanning tree \citep{CMG}. Its main focus is on problems arising in
imaging applications. The spanning tree construction and clustering presented
in CMG do not lend themselves to a simple parallel implementation.

Lean Algebraic Multigrid uses a more standard AMG approach with modifications
suited for Laplacian matrices. Notably, it employs unsmoothed aggregation
tailored to scale-free graphs, a specialized distance function, and a Krylov
method to accelerate solutions on each of the multigrid levels. These changes
are not rooted in graph theory but produce good empirical results. Empirically, LAMG is
slightly slower than CMG but more robust \citep{LAMG}. LAMG's partial
elimination procedure and clustering process are both inherently serial.

Degree-aware Rooted Aggregation applies a similar partial elimination technique to LAMG, except it is limited to degree 1 vertices.
Its performance relies on a combination of unsmoothed aggregation based on vertex degree and a multilevel Krylov method called K-cycles\citep{DRA}.
A K-cycle is a multigrid W-cycle (one can imagine a K-cycle with different cycle index, but we only consider K-cycles with a cycle index of 2) with Krylov acceleration applied at each level.
DRA's aggregation (like LAMG's) is inherently serial and would require modifications for parallelism.

\subsection{Algebraic Multigrid}

The three solvers mentioned previously have one thing in common: they use some sort of algebraic multigrid method.
Algebraic multigrid (AMG) is a family of techniques for solving linear systems of the form $Lx=b$.
Specifically, algebraic multigrid constructs a hierarchy of approximations to $L$ using only the values in the matrix $L$.
These approximations, $\{L=L_0, L_1, L_2, L_3, ...\}$, are successively coarser: $\text{size}(L_l) > \text{size}(L_{l+1})$.
For each level $l$ in the hierarchy, we have a restriction operator $R_l$ that transfers a residual from level $l$ to level $l+1$ and a prolongation $P_l$ that transfers a solution from level $l+1$ to level $l$.

\begin{algorithm}
  \caption{Multigrid cycle with cycle index $\gamma$}
    \label{alg:vcycle}
  \begin{algorithmic}[1]
    \Function{mgcycle}{level $l$, initial guess $x$, rhs $b$}
      \If{$l$ is the coarsest level}
        \Let{$x$}{Direct solve on $L_{l} x = b$}
        \State \Return{$x$}
      \Else
        \Let{$x$}{smooth($x, b$)} \Comment{Pre-smoothing}
        \Let{$r$}{$b - L_l x$} \Comment{Residual}
        \Let{$r_c$}{$R_l r$} \Comment{Restriction}
        \Let{$x_c$}{$0$}
        \For{$i \in [0,\gamma)$}
          \Let{$x_c$}{\Call{mgcycle}{$l+1, x_c, r_c$}} \Comment{Coarse level solve}
        \EndFor
        \Let{$x$}{$x + P_l x_c$} \Comment{Prolongation}
        \Let{$x$}{smooth($x, b$)} \Comment{Post-smoothing}
        \State \Return{$x$}
      \EndIf
    \EndFunction
  \end{algorithmic}
\end{algorithm}

Algorithm \ref{alg:vcycle} depicts a multigrid cycle with cycle index $\gamma$ ($\gamma=1$ is called a V-cycle, and $\gamma=2$ is a W-cycle). Repeated application of this algorithm $x^{k+1} \gets \textsc{mgcycle}(0, x^k, b)$ often converges to a solution $L x^* = b$, reaching a given tolerance in a number of iterations that is bounded independent of problem size. Fast convergence depends on sufficiently accurate restriction and prolongation operators complemented by pre-smoothing and post-smoothing that provide local relaxation. Typical smoother choices are Gauss-Seidel, Jacobi, and Chebyshev iteration.
Note that pre- and post-smoothing may use different smoothers or different numbers of smoothing iterations.
Given a restriction and prolongation, we identify ``low frequencies'' as those functions that can be accurately transferred to a coarse space and back.
A smoother need only be stable on such functions but must reduce the error uniformly for all ``high frequencies''---those which cannot be accurately transferred.

There are two main ways of constructing $\{L_1, L_2, L_3, ...\}$: classical AMG and aggregation-based AMG. Classical (or Ruge-St{\"u}ben) AMG constructs $R$ and $P$ using a coarse-fine splitting: the coarse grid is a subset of the degrees of freedom (or ``points'') of the fine grid \citep{rugestuben}. $P$ keeps values at coarse points and extends them via a partition of unity to neighboring fine points. Then, $R=P^T$. In aggregation-based AMG, degrees of freedom are clustered into aggregates. An aggregate on the fine level becomes a point on the coarse level. $R$ and $P$ usually take some weighted average of points in each aggregate to the coarse level. Aggregation-based multigrid methods typically smooth $R$ and $P$ for better performance~\citep{vanek1996algebraic}. In both classical and aggregation-based AMG, the coarse level matrix is usually constructed via a Galerkin product: $L_{l+1} = R_l L_l P_l$.

\subsubsection{Measuring Performance}\label{sec:wda}

In order to evaluate the performance of our solver relative to the state of the art, we need some performance metric. We could use runtime, but it is dependent on the implementation of the algorithms (for example, a MATLAB implementation would probably be slower than a C++ implementation). Instead, we will use work per digit of accuracy (WDA) to measure performance.
\begin{align*}
\text{WDA} &= \frac{-\text{work}}{\log_{10} \Delta r}, \quad r = b-Lx, \quad \Delta r = \frac{\lVert r_{\text{final}}\rVert}{\lVert r_{\text{initial}} \rVert} \\
\text{work} &= \frac{\text{total FLOPS}}{\text{FLOPS to compute residual on finest level}}
\end{align*}

WDA measures how much work is required to reduce the residual by an order of magnitude.
$r$ is the residual norm, and $\Delta r$ is the change in residual norm from an initial solution to a final solution.
Work is expressed in terms of the number of FLOPS required for a solve divided by that required to compute a residual on the finest level.
This measure is also proportional to required memory transfers and is typically approximated in terms of the number of nonzeros in the sparse operators $L_l, R_l, P_l$.
WDA only measures the efficiency of a single solve; it does not take the setup phase into account.
WDA also does not account for parallel scalability.

\subsubsection{AMG as a Preconditioner}\label{sec:precond}

AMG can be used as a solver by itself, but it often provides faster and more robust convergence when used as a preconditioner for a Krylov method. In order to solve $Lx=b$, the Krylov method is applied to $MLx=Mb$, where $M$ is a single AMG cycle, such as a V-cycle or W-cycle.  The convergence of the Krylov method will depend on the spectrum of $ML$, converging in a number of iterations bounded by square root of the condition number.
We use our multigrid solver as a preconditioner to conjugate gradients, or in the case of K-cycles, flexible CG (see \S\ref{sec:kcycle}).

\section{Main Contribution}

Our solver uses an unsmoothed aggregation based multigrid technique with low degree elimination. Its notable features are: 1. a matrix distribution that improves parallel performance but increases complexity of aggregation and elimination algorithms 2. a parallel elimination algorithm using this matrix distribution 3. a parallel aggregation algorithm also using this matrix distribution. We started building our solver by analyzing the performance of LAMG and its potential for parallelism.

\subsection{Issues With a Parallel Implementation of LAMG}\label{sec:lamgissues}

To understand what parts of LAMG we could adapt to a distributed
memory setting, we ran LAMG on 110 graphs from the University of Florida Sparse
Matrix Collection \citep{SPCol} while varying its parameters for cycle index,
smoother and iterate recombination (we would like to also vary the aggregation
routine, but LAMG's energy \textit{unaware} aggregation is not working in the
MATLAB implementation). LAMG uses a cycle index of 1.5, which is half way between a V-cycle and a W-cycle. Iterate recombination is a multilevel Krylov method that chooses the optimal search direction from the previous solution guesses at each level. It is similar in nature to K-cycles (see section~\ref{sec:kcycle}). Our goal was to find which parts of LAMG had the
largest effect on solver performance.

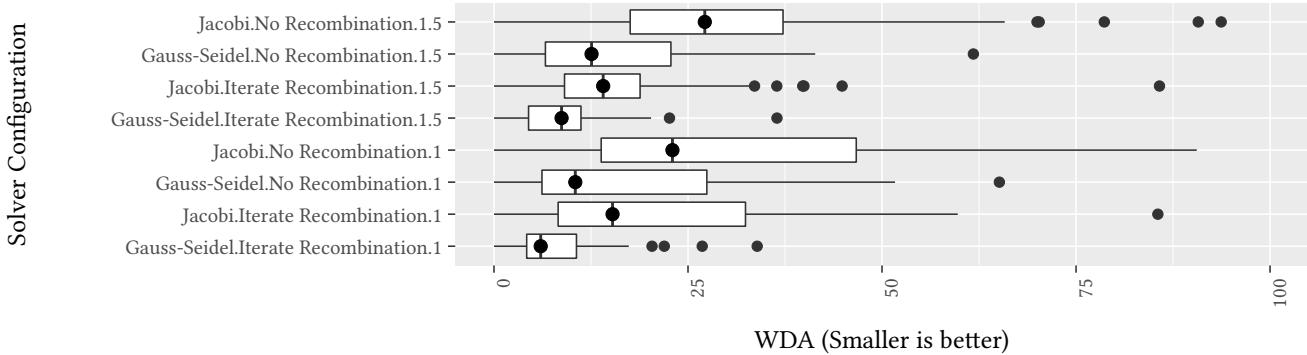
\begin{figure*}
  \centering
  \resizebox{\linewidth}{!}{\input{serial_box_lamg}}
  \vspace{-2.5em}
  \caption{
    Boxplots of the performance of serial LAMG \citep{LAMGDL} on 110 graphs from the University of Florida Sparse Matrix collection \citep{SPCol}.
    Solver configuration (vertical axis) is a triple of smoother, iterate recombination (or not), and cycle index.
    Performance is measured in terms of work per digit of accuracy (see section~\ref{sec:wda}).
    WDA accounts for work per iteration and number of iterations.
    Problems are solved to a relative tolerance of $10^{-8}$.
    Each box represents the interquartile range of WDA for a given solver configuration.
    The line and dot inside the box indicates the median WDA.
    Horizontal lines on either side of the box indicate the range of WDA values.
    Dots outside the box indicate outliers.
  }
  \label{fig:serial_box_lamg}
\end{figure*}

Figure~\ref{fig:serial_box_lamg} shows the result of these tests. Visually,
Gauss Seidel smoothing and iterate recombination together give the
smallest WDA. Cycle index has a very small effect relative to smoother and
iterate recombination.

\begin{figure}
  \centering
  \includegraphics[width=\columnwidth]{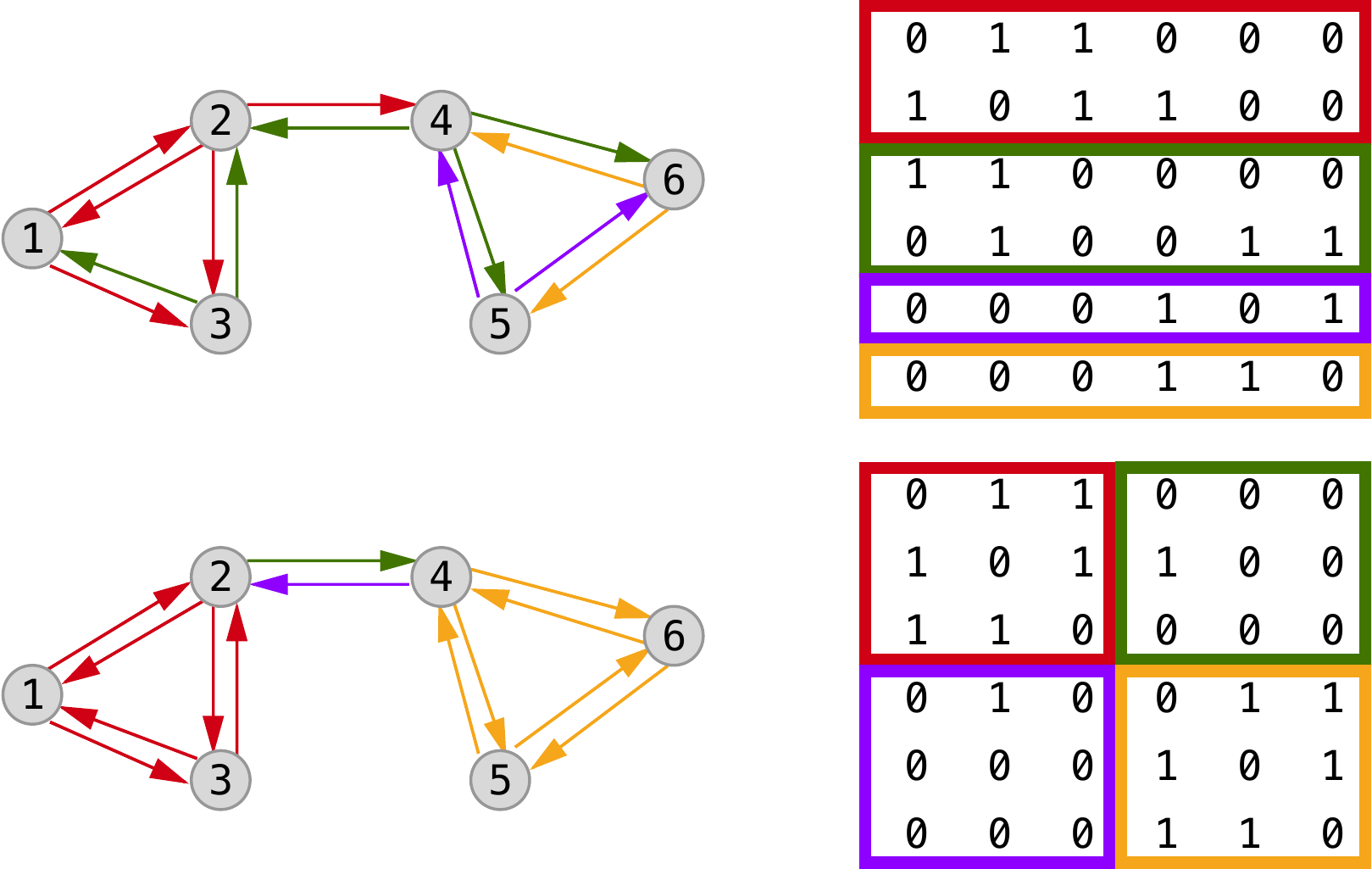}
  \caption{The distribution of directed edges in the graph (left) and adjacency
  matrix (right). The top is a 1D vertex distribution, and the bottom is a 2D edge
  distribution. Each color corresponds to edges and matrix entries owned by
  processor. Note that no process in the 2D distribution has all the out-edges
  or in-edges for a given process.}
  \label{fig:edge_dist}
\end{figure}

We applied a statistical analysis to LAMG's performance with its features on and off.
The largest change in solver performance came from using Gauss-Seidel smoothing over Jacobi smoothing.
On social network graphs, Gauss-Seidel smoothing often performs much better than Jacobi smoothing.
Iterate recombination is the second most important factor.
It helps improve performance on high diameter graphs such as road networks.
The least important factor is the 1.5 cycle index.
Although it does slightly improve WDA, it is much less important than Gauss-Seidel smoothing and iterate recombination.

Given the importance of Gauss-Seidel smoothing and iterate recombination, we would like to use them in a parallel implementation. However, there are a couple of challenges with using these features in a parallel implementation of LAMG:

\begin{enumerate}
  \item The power-law vertex degree distribution can cause large work and communication imbalances.
  \item LAMG's low degree elimination is a sequential algorithm.
  \item LAMG's energy-based aggregation is a sequential algorithm.
  \item Multilevel Krylov acceleration can be a parallel bottleneck.
  \item W-cycles exhibit poor parallel performance.
  \item Gauss-Seidel smoothing is infeasible in parallel.
\end{enumerate}

\subsection{2D Matrix Distribution}

Our initial implementation used a vertex distribution of the graph: each processor owns some number of rows in the Laplacian matrix.
This initial implementation scaled very poorly as we increased the number of processes.
One process had many more edges than all the other processes, causing a work imbalance.
Using a partitioner can temporarily alleviate this problem, but in the limiting case (where each process owns a single vertex), processes with hubs will have significantly more work than those without.
Research on scaling sparse matrix operations on adjacency matrices suggests that a more sophisticated distribution of matrix entries is important.
We use CombBLAS, which has demonstrated the scalability and load balancing benefits of a 2D matrix distribution~\citep{CombBLAS}.
A 2D matrix distribution can be thought of as a partition of graph edges instead of vertices.
Computational nodes (or processes) form a 2D grid over the matrix (called a processor grid).
Each process is given a block of the matrix corresponding to its position in the grid (see Figure~\ref{fig:edge_dist}).
Vectors can either be distributed across all processes or just processes on the diagonal.
In our implementation, we found performance did not change with vector distribution, so we distribute them across diagonal entries in the processor grid.
The disadvantage of a 2D distribution is that it has higher constant factors and poorer data locality.

CombBLAS expresses graph algorithms in the language of linear algebra. Instead
of the usual multiplication and addition used in matrix products, CombBLAS allows the user to use
custom multiplication and addition operations. We will follow the $\oplus.\otimes$ notation used by GraphBLAS (the standardization effort of CombBLAS)\citep{graphblas}. Here, $\oplus$ specifies the custom addition operator and $\otimes$ the multiplication.
While an $m\times n$ matrix $A$ maps from $\mathbb R^n$ to $\mathbb R^m$, a generalized matrix $\hat A$ maps from $\mathbb C^n$ to $\mathbb D^m$, where $\mathbb C$ and $\mathbb D$ may be different.
\begin{align*}
  (A v)_i \coloneqq \sum_{j} A_{ij} v_j && \text{Usual matrix product} \\
  (\hat A \: {\oplus.\otimes} \: v)_i \coloneqq \bigoplus_j \hat A_{ij} \otimes v_j && \text{Generalized matrix product}
\end{align*}
\begin{align*}
  \hat A \in \mathbb{B}^{m \times n},& \quad v \in \mathbb{C}^n \\
  \otimes \colon \mathbb{B} \times \mathbb{C} \rightarrow \mathbb{D},& \quad \oplus \colon \mathbb{D} \times \mathbb{D} \rightarrow \mathbb{D}
\end{align*}

Our operators are similar in structure to a semiring, but they permit different types of elements in the input matrix and the input and output vectors.
If we structure our algorithms in terms of generalized matrix-vector products, we can piggyback on the proven performance of CombBLAS.
However, if we cannot express all parts of an algorithm using this linear algebraic approach, we must keep in mind that no computational node has a complete view of all the edges to and from any vertex.
Implementing an arbitrary vertex neighborhood operation, such as choosing the median of neighbors, would require potentially non-scalable custom communication.

\subsection{Random Vertex Ordering}

A 2D matrix distribution alleviates communication bottlenecks and some load balancing difficulties, but
the processes responsible for diagonal blocks are often found to have many more nonzeros than typical off-diagonal blocks.
For example, social network often have a couple of large degree ``hubs'' that are
connected to many other vertices. These vertices correspond to an almost dense
column and row in the graph Laplacian. Often a single process will end up with a
few hubs and have 10x (or more) edges than other processes. A simple
technique to better balance the workload is to randomly order vertices. This
trades data locality for better load distribution. More sophisticated
techniques exist for 2D matrix partitioning \citep{2d-part}, but we found that
a random distribution is sufficient for acceptable load balance. We found that
random vertex ordering increased not only asymptotic parallel scalability but also
performance for relatively small process counts. We apply this randomization only to the input matrix; we do not re-randomize at coarser levels.

\subsection{Parallel Low-Degree Elimination}

Low-degree elimination greatly reduces problem complexity in graph Laplacians,
especially those arising from social networks. Like LAMG, we eliminate vertices
of degree 4 or less.

The main difficulty in adapting low-degree elimination to a distributed memory
system is deciding which vertices to eliminate. If we had a vertex centric
distribution, each process could locally decide which of its local vertices to
eliminate. However, we have a 2D edge distribution, so we will instead structure
our elimination in terms of linear algebra and allow CombBLAS to do the heavy
lifting.

Our algorithm for low-degree elimination is detailed in
Algorithm~\ref{alg:elim}. It essentially boils down to two steps. First, mark
all vertices of degree 4 or less as candidates for elimination. Then, for each
candidate, check whether it has the lowest hash value among all neighboring
candidates. If it has the lowest hash value, it will be eliminated. The hash
value is a hash of the vertex's id. We use a hash of the id instead of
the id itself in order to prevent biases that might occur when using a non-random matrix
ordering.

In linear algebraic terms, choosing vertices to eliminate can be accomplished by creating a vector that
marks each vertex as a candidate or not (line \ref{elim:line:mark} of Algorithm~\ref{alg:elim})
and then multiplying said vector with the Laplacian matrix using a custom
$\otimes$ and $\oplus$ (line \ref{elim:line:mark}). The $\otimes_{\text{elim}}$ filters out matrix entries that
are not candidates and neighbors:
$$
  a \otimes_{\text{elim}} c = \begin{cases}
    c & \text{if } a \ne 0 \\
    \varnothing & \text{otherwise}
  \end{cases}
$$
where $c = \varnothing$ is used to indicate that a vertex should not be considered.

The $\oplus_{\text{elim}}$ chooses the candidate with the smallest
hashed id,
$$
  x \oplus_{\text{elim}} y = \begin{cases}
    x & \text{if } \text{ hash}(x) \le \text{hash}(y)\\
    y & \text{if } \text{ hash}(x) > \text{hash}(y)
  \end{cases}
$$
where $\text{hash}(\varnothing) = \infty$.

\begin{algorithm}[h]
  \caption{Determine vertices to eliminate}
    \label{alg:elim}
  \begin{algorithmic}[1]
    \Function{Low-Degree Elimination}{$L\in \mathbb{R}^{n \times n}$}
      \Let{$\mathit{candidates}_i$}{i \textbf{if} degree$(V_i) \le 4$ \textbf{else} $\varnothing$, \quad $i \in V$}\label{elim:line:mark}
      \Let{$z$}{$L \: {\oplus_{\text{elim}}.\otimes_{\text{elim}}} \: \mathit{candidates}$}\label{elim:line:filter}
      \State \textbf{if} $z_i = i$ \textbf{then} eliminate $V_i$ \label{line:elimination}
    \EndFunction
  \end{algorithmic}
\end{algorithm}

We use the Laplacian $L$ so that the neighborhood of each vertex contains itself. Each entry $z_i$ corresponds to the neighbor of $v_i$ that is a candidate and has the lowest hashed id. If $z_i = i$, then we know that $v_i$ is the candidate with the smallest hash among its neighbors (line~\ref{line:elimination}) and can be eliminated. Let $\mathcal{F}$ denote eliminated vertices and $\mathcal{C}$ be vertices that have not been eliminated. Following Livne and Brandt \citep{LAMG}, we express elimination in the language of multigrid.
We first introduce a permutation $\Pi$ of the degrees of freedom such that
\begin{equation*}
  L_l = \Pi \begin{pmatrix*}
    L_{\mathcal{F}\mathcal{F}} & L_{\mathcal{F}\mathcal{C}} \\
    L_{\mathcal{F}\mathcal{C}}^T & L_{\mathcal{C}\mathcal{C}}
  \end{pmatrix*} \Pi^T
\end{equation*}
which admits the block factorization
\begin{equation*}
  L_l = \Pi
  \begin{pmatrix*} I & \\ L_{\mathcal{FC}}^T L_{\mathcal{FF}}^{-1} & I \end{pmatrix*}
  \begin{pmatrix*} L_{\mathcal{FF}} & \\ & L_{l+1} \end{pmatrix*}
  \begin{pmatrix*} I & L_{\mathcal{FF}}^{-1} L_{\mathcal{FC}} \\ 0 & I \end{pmatrix*}
    \Pi^T \label{eq:blockldu}
\end{equation*}
in terms of the Schur complement
\begin{equation*}
  L_{l+1} = L_{\mathcal{CC}} - L_{\mathcal{FC}}^T L_{\mathcal{FF}}^{-1} L_{\mathcal{FC}} .
\end{equation*}
Note that $L_{\mathcal{FF}}$ is diagonal so its inverse is also diagonal and that
we can alternately express $L_{l+1} = P_l^T L_l P_l$ in terms of the prolongation
\begin{equation*}
  P_l = \Pi \begin{pmatrix*}
    -L_{\mathcal{FF}}^{-1} L_{\mathcal{FC}} \\
    I
  \end{pmatrix*}.
\end{equation*}
Inverting the block factorization yields
\begin{equation*}
  \begin{split}
    L_l^{-1}
    &= \Pi \begin{pmatrix*} I & - L_{\mathcal{FF}}^{-1} L_{\mathcal{FC}} \\ 0 & I \end{pmatrix*}
    \begin{pmatrix*} L_{\mathcal{FF}}^{-1} & \\ & L_{l+1}^{-1} \end{pmatrix*}
    \begin{pmatrix*} I & \\ -L_{\mathcal{FC}}^T L_{\mathcal{FF}}^{-1} & I \end{pmatrix*} \Pi^T \\
    &= P L_{l+1}^{-1} P^T + \Pi \begin{pmatrix*} L_{\mathcal{FF}}^{-1} & 0 \\ 0 & 0 \end{pmatrix*} \Pi^T
  \end{split}
\end{equation*}
which is an additive 2-level method with $\mathcal{F}$-point smoother
\begin{equation*}
  \mathrm{\mathcal{F}smooth}(x,b) = \Pi \begin{pmatrix*} L_{\mathcal{FF}}^{-1} & 0 \\ 0 & 0 \end{pmatrix*} \Pi^T b.
\end{equation*}
Rather than applying $L_{l+1}^{-1}$ exactly, we approximate it by continuing the multigrid cycle.

The ids of eliminated vertices are broadcast down processor rows and columns. Each process constructs entries of $L_{\mathcal{F}\mathcal{C}}$ and $L_{\mathcal{F}\mathcal{F}}^{-1}$ that depend on its local entries in $L_l$. These constructed entries are then scattered to the processes that own them in $P_l$. Alternatively, we could construct $\Pi$, $L_{\mathcal{F}\mathcal{C}}$, and $L_{\mathcal{F}\mathcal{F}}^{-1}$ explicitly and use them to build $P_l$.

Our candidate selection scheme is not as powerful as the serial LAMG scheme.
The serial scheme will eliminate every other vertex of a chain. In the best
case we do the same, but in the worst case we eliminate only one vertex if the
hash values of vertices in the chain are in sequential order.
To address this issue, we can run low-degree elimination multiple times in a row to eliminate more of the graph.
In practice, we find one iteration is sufficient to remove most of the low degree structure.

We apply low-degree elimination before every aggregation level and only if more than 5\% of the vertices will be eliminated.

\newcommand\status{\mathit{status}}
\newcommand\votes{\mathit{votes}}
\newcommand\aggs{\mathit{aggregates}}
\newcommand\loc{\mathit{local\_votes}}

\subsection{Parallel Aggregation}

\begin{algorithm}
  \caption{Aggregation}
    \label{alg:agg}
  \begin{algorithmic}[1]
    \Function{Aggregation}{$S \in \mathbb{R}^{n \times n}$}
      \Let{$\status_i$}{$(\mathit{Undecided}, i)$, \textbf{for} $i$ \textbf{in} $1..n$}
      \Let{$\votes_i$}{$0$, \textbf{for} $i$ \textbf{in} $1..n$}
      \For{$\mathit{iter}$ \textbf{in} $1..10$}
        \State{$\status, \votes$}
        \Statex[3] $\gets$ \Call{Aggregation-Step}{$S, \status, \votes, 0.5^{\textit{iter}}$}
      \EndFor
      \For{$i$ \textbf{in} $1..n$}
        \Let{$(\cdot,j)$}{$\status_i$}
        \Let{$\aggs_i$}{$j$}
      \EndFor
      \State \Return $\aggs$
    \EndFunction

    \Statex[0]\Comment{$\status$ is a vector with elements of type $(\mathit{State}, \mathit{Index})$}
    \Statex[0]\Comment{$S$ is the strength of connection matrix}
    \Function{Aggregation-Step}{$S, \status, \votes, \textit{filter-factor}$}
      \Let{$S_{\textit{filt}}$}{Remove nonzeros $<$ \textit{filter-factor} from $S$}
      \Let{$d$}{$S_{\textit{filt}} \: {\oplus_{\text{agg}}.\otimes_{\text{agg}}} \: \status$}
        \Let{$\loc$}{Sparse map containing votes for vertices}
        \For{$i$ \textbf{in} $1..n$}
          \Let{$(s, j, w)$}{$d_i$}
          \If{$s = \mathit{Seed}$}
            \Statex\Comment{Found a neighboring seed,}
            \Statex\Comment{$V_i$ is aggregated with $V_n$}
            \Let{$\status_i$}{$(\mathit{Decided},j)$}
          \ElsIf{$s = \mathit{Undecided}$}
            \Statex\Comment{No neighboring seed, $V_i$ votes for $V_n$}
            \Let{$\loc[j]$}{$\loc[j]+1$}
          \EndIf
        \EndFor

        \Statex[1]\Comment{Communicate local votes}
        \Let{$\loc$}{$\text{reduce\_by\_key}(+, \loc)$}
        \Statex[1]\Comment{Update persistent votes counts}
        \Let{$\votes$}{$\votes + \loc$}

        \For{$i$ \textbf{in} $1..n$}
          \If{$\votes_i > 8 \amp status_i = (\mathit{Undecided},i)$}
            \Statex\Comment{Vertices with enough votes become $\mathit{Seed}$s}
            \Let{$\status_i$}{$(\mathit{Seed}, i)$}
          \EndIf
        \EndFor

        \State \Return $\status, \votes, \aggs$
    \EndFunction
  \end{algorithmic}
\end{algorithm}

Our parallel aggregation algorithm (Algorithm~\ref{alg:agg}) uses a strength of connection metric, $S$, to determine how to form aggregates.
It indirectly determines how likely any two vertices will be clustered together.
We use the affinity strength of connection metric proposed by Livne and Brandt in the LAMG paper \citep{LAMG}.
To construct the strength of connection matrix, $S$, we smooth four vectors three times each.
The relevant parts of the smoothed vectors are broadcast down communication grid rows and columns.
Each process then constructs its local part of the matrix using the affinity metric
\begin{align*}
  L &\in \mathbb{R}^{n \times n}, y \in \mathbb{R}^{n \times m}, m \ll n, \text{random entries} \\
  x &\coloneqq \text{smooth on } Ly=0 \\
  C_{ij} &\coloneqq
    \begin{dcases*}
      0 & if $A_{ij} = 0$ or $i = j$ \\
      \frac{|\sum_{k=1}^m x_{ik} x_{jk}|^2}
        {(\sum_{k=1}^m x_{ik} x_{ik})^2 (\sum_{k=1}^m x_{jk} x_{jk})^2} & otherwise
    \end{dcases*} \\
\end{align*}
\begin{align*}
  S_{ij} &\coloneqq \frac{C_{ij}}{\max{(\max_{s \neq i} C_{is}, \max_{s \neq j} C_{s,j})}}
\end{align*}
where $A$ is the adjacency matrix.
The smoothing we use is three iterations of Jacobi smoothing.
The total cost of creating $S$ is 4 vectors $*$ 3 smoothing iterations, for 12 matrix-vectors multiplies total.
Currently, we smooth each vector separately and do not exploit any of the parallelism available in smoothing multiple vectors together.

The construction of $C_{ij}$ is entirely local because $C$ has the same distribution as $A$. Constructing $S$ requires communication along processor rows and columns to find the largest nonzero of $C$ for each column and row in the matrix. Note that $S$ has $0$ diagonal.

Our aggregation algorithm uses a voting scheme in which each vertex votes for which of its neighbors it would like to aggregate with.
A vector $status$ contains a state $\in \{\mathit{Seed}, \mathit{Undecided}, \mathit{Decided}\}$ and an index for each vertex,
$$
  \status_i \coloneqq
    \begin{dcases*}
      (\mathit{Seed}, i) & Vertex $i$ is a $\mathit{Seed}$\\
      (\mathit{Undecided}, i) & Vertex $i$ has not yet joined an aggregate \\
      (\mathit{Decided}, j) & Vertex $i$ is aggregated with $\mathit{Seed}$ $j$ \\
    \end{dcases*}
$$

Initially, $\status_i = (\mathit{Undecided}, i)$.
In each voting iteration, each $\mathit{Undecided}$ vertex either aggregates with a neighboring $\mathit{Seed}$ (and becomes $\mathit{Decided}$) or votes for a neighboring $\mathit{Undecided}$ vertex to become a $\mathit{Seed}$.
If a vertex is voted for enough times, it will turn into a $\mathit{Seed}$.
The strength of connection matrix determines which neighboring vertex is aggregated to or voted for.
For the first round of aggregation, we only consider very strong connections ($> 0.5$).
We gradually reduce this bound in subsequent iterations (Livne and Brandt apply a similar technique in LAMG~\citep{LAMG}).
The vertices' choice of neighbor is expressed as a matrix-vector product $S \: {\oplus_{\text{agg}}.\otimes_{\text{agg}}} \: \status$ with
\begin{align*}
  w \otimes_{\text{agg}} (\mathit{state},i) &\coloneqq
    \begin{cases}
      (\mathit{state},i,w) & \text{if } w \neq 0 \\
      (\mathit{Decided}, -1, 0) & \text{otherwise} \\
    \end{cases}
\end{align*}
\begin{align*}
  (\mathit{state}_a, i_a, w_a) &\oplus_{\text{agg}} (\mathit{state}_b, i_b, w_b) \coloneqq \\
    &\begin{cases}
      (\mathit{state}_a, i_a, w_a) & \text{if } \mathit{state}_a = \mathit{state}_b \amp w_a \ge w_b \\
      (\mathit{state}_b, i_b, w_b) & \text{if } \mathit{state}_a = \mathit{state}_b \amp w_a < w_b \\
      (\mathit{state}_a, i_a, w_a) & \text{if } \mathit{state}_a > \mathit{state}_b \\
      (\mathit{state}_b, i_b, w_b) & \text{if } \mathit{state}_a < \mathit{state}_b \\
    \end{cases}
\end{align*}
where $\mathit{Seed} > \mathit{Undecided} > \mathit{Decided}$.
Note that the input vector contains pairs, whereas the output vector contains 3-tuples.

The votes for each vertex are tallied using a sparse reduction that has the same communication structure as a matrix-vector product.
Our implementation uses an allreduce because it has lower constants and is not a bottleneck.
Using the tallied votes, we update the status vector with new roots and aggregates.
The vote counts are persisted across voting iterations so that vertices with low degree may eventually become seeds.
We choose to do 10 voting iterations, and we convert $\mathit{Undecided}$ vertices to $\mathit{Seed}$s if they receive 8 or more votes.
Both these numbers are arbitrary.
In practice, we find that performance is not sensitive to increasing or decreasing these constants by moderate amounts.

The result of aggregation is a distributed vector $v$, where vertex $i$ is part of aggregate $v_i$.
We perform a global reordering so that aggregates are numbered starting at $0$.
We construct $R$ by inserting $R_{v_j j}=1$, where $j$ is in the locally owned portion of $v$, then scattering to a balanced 2D distribution.
This 2D distribution is similar to $L$, except each process has a rectangular local block instead of a square one.
\begin{align*}
  R_{ij} &\coloneqq \begin{cases*}
    1 & if $v_j = i$ ($V_j$ is in aggregate $i$) \\
    0 & otherwise
  \end{cases*} \\
  P &\coloneqq R^T, \quad L_{l+1} \coloneqq R_l L_l P_l
\end{align*}

\subsection{Smoothing}\label{sec:smoothing}

In general, Gauss-Seidel smoothing is the best performing smoother on graph
Laplacians (section~\ref{sec:lamgissues} provides more details on Gauss-Seidel
vs Jacobi performance). However, its parallel performance on graph Laplacians
is very poor. Most processes have an overwhelming amount of connections that
reference values outside of the local block of the matrix, and the graph cannot be colored with a reasonable number of colors. Our resulting choice
of smoother is Chebyshev/Jacobi smoothing~\citep{Adams-02} because it is stronger than
(weighted) Jacobi with equivalent parallel performance. Instead of applying $k$ iterations of a smoother, we use one application of degree $k$ Chebyshev smoothing. We choose our lower and upper bounds because $.3$ and $1.1$ times the largest eigenvalue as estimated by $10$ Arnoldi iterations (we include these iterations in our setup cost).

\subsection{K-cycles}\label{sec:kcycle}

We would like to include some form of multilevel Krylov acceleration because it
improves solver robustness (see section~\ref{sec:lamgissues} for more details).
We implemented K-cycles as described by Notay and Vassilevski in ``Recursive
Krylov-based multigrid cycles'' \citep{notay2008recursive} and used in Napov and
Notay's DRA \citep{DRA}. At each aggregation level in our multigrid hierarchy,
we perform a number of Flexible Conjugate Gradient (FCG) \citep{notay2000flexible} iterations using the rest of the hierarchy as a
preconditioner. We do not apply Krylov acceleration to elimination levels because
they exactly interpolate the solution from the coarse grid.

\section{Numerical Results}\label{sec:num}

Our solver uses a V-cycle or K-cycle with one iteration of degree 2 Chebyshev smoothing before
restriction and one iteration of degree 2 Chebyshev smoothing after prolongation. The V-cycle is
used as a preconditioner for Conjugate Gradient (FCG when using K-cycles). Our solver is implemented in
C++ and uses CombBLAS \citep{CombBLAS} for sparse linear algebra and PETSc
\citep{petsc-user-ref,petsc-efficient} for Chebyshev smoothing, eigenvalue estimation, and Krylov methods.

Our numerical tests were run on NERSC's Edison and Cori clusters.
Edison is a Cray XC30 supercomputer with 24 ``Ivy Bridge'' Intel Xeon E5-2695 v2 cores per node and a Cray Aries interconnect.
Cori is Cray supercomputer with 36 ``Haswell'' Intel Xeon E5-2698 v3 cores per node and a Cray Aries interconnect.
For each test, we run four MPI processes per physical node in order to obtain close to peak bandwidth.

We solve to a relative tolerance of $10^{-8}$.
We use a random right hand side with the constant vector projected out.
We also tested with a right hand side composed of low eigenmodes but did not notice any difference in performance compared to a random right hand side.
The coarsest level size is set to have no more than 1000 nonzeros in $L$.
Our solver uses somewhere from 20 to 40 levels depending on the problem. In order to ensure enough work per process, we use a smaller number of processors on coarser levels if the amount of work drops below a threshold (if $\text{nnz}(L)/10000 < $ number of processors).
The coarsest level always ends up on a single process.

\subsection{Comparison to Serial}\label{sec:comparelamg}

\begin{figure*}
  \centering
  \resizebox{\linewidth}{!}{\input{serial_box}}
  \vspace{-2.5em}
  \caption{
    Boxplots of solver performance in various configurations on a selection of 110 graphs from the University of Florida Sparse Matrix Collection \citep{SPCol}.
    Performance is measured in terms of work per digit of accuracy (see section~\ref{sec:wda}).
    WDA accounts for work per iteration and number of iterations.
    The last number in the solver configuration indicates the cycle index. All solves to a relative tolerance of $10^{-8}$.
    LAMG with Jacobi smoothing, no recombination, and cycle index 1 has many undisplayed outliers because they fall well above 100 WDA.
  }
  \label{fig:serial_box}
\end{figure*}
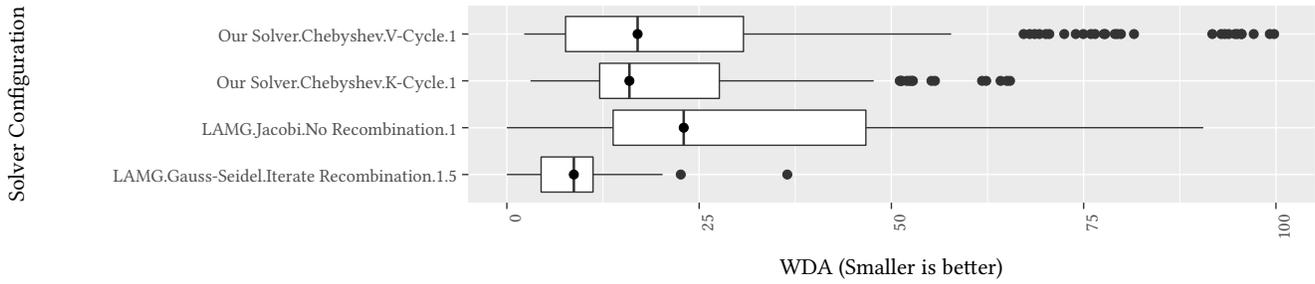

We compared our parallel solver to Livne and Brandt's serial LAMG implementation
in MATLAB \citep{LAMG,LAMGDL}. It is hard to fairly compare
single threaded performance between our solver and LAMG. One or the other could be
better optimized, or choice of programming language might make a difference. To
provide a fair comparison, we measure the work per digit of accuracy of each
solver.

In general, our method exhibited poorer convergence factors than the serial
LAMG implementation. We expect that our solver performs worse because we have made
multiple concessions for parallel scalability. These changes are:

\begin{enumerate}
  \item No energy-based aggregation
  \item Chebyshev smoothing (versus Gauss-Seidel)
  \item No 1.5 cycle index
  \item No multilevel Krylov acceleration (for V-cycles)
\end{enumerate}

Figure~\ref{fig:serial_box} shows the performance (measured in terms of WDA) of serial LAMG and our solver.
The fourth line shows LAMG with all of its parallel-unfriendly features enabled.
The third line shows LAMG without these features (but using LAMG's standard aggregation and elimination).
The first line is our solver without K-cycles and the second line with K-cycles.
Our solver has a higher median WDA and variance than LAMG with all features enabled.
Our solver is not as robust and has more outliers.
Most of these outliers are road networks.
A couple of graphs have a fairly high WDA with our solver but are solved quickly by LAMG.
This is expected because we have made concessions in order to achieve parallel performance.
However, our solver performs much better than LAMG with parallel friendly features (Jacobi smoothing and no recombination).

\begin{figure}
  \centering
  \resizebox{\columnwidth}{!}{\input{kcycles}}
  \vspace{-2em}
  \caption{
    Loglog plot of strong scaling of our solver with K-cycles and with V-cycles on the \textit{hollywood} graph (1,139,905 vertices, 113,891,327 edges) on Edison.
    Numeric labels next to points indicate number of processes for a given solve.
    There are 21-23 multigrid levels so the coarsest level is visited $2^{10}-2^{11}$ times with an index 2 cycle.
    The coarse level solves and redistribution become a parallel bottleneck.
  }
  \label{fig:kcycles}
\end{figure}
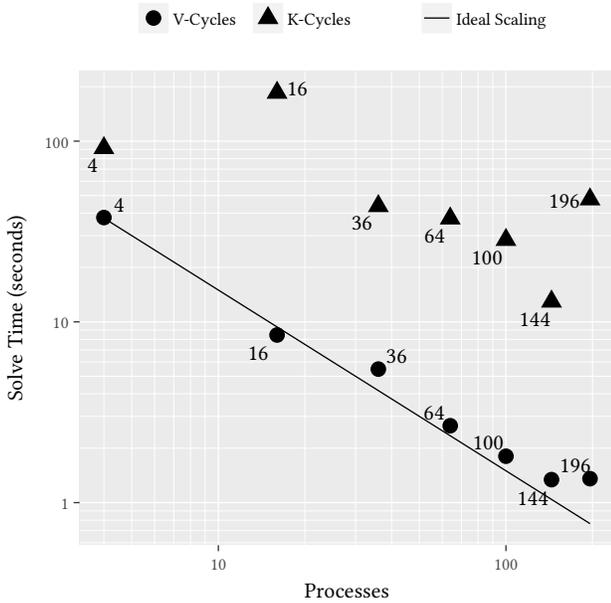

Also interesting to note is the small difference in WDA of our solver with
and without K-cycles. K-cycles have low variance, and hence are more robust, but median performance is not much improved.
However, this small gain in WDA is not worth the high parallel cost of K-cycles (as seen in Figure~\ref{fig:kcycles}).
On the \textit{hollywood} graph, K-cycles are clearly slower and scale worse than V-cycles.
K-cycles need inner products on every level of the cycle (as with Krylov smoothers), which require more communication.
The W-cycle structure of K-cycles also causes a parallel bottleneck.
The coarsest level is visited $2^{10}-2^{11}$ times, which results in lots of sequential solves and data redistributions.
Because K-cycles appear worse than V-cycles (especially when compared to the marginal robustness they provide), we use V-cycles by default in our solver.
All following performance results for our solver use V-cycles.

\subsection{Strong Scaling}

\begin{figure}
  \centering
  \resizebox{\columnwidth}{!}{\input{efficiency}}
  \vspace{-2em}
  \caption{
    Semilog-x plot of efficiency $\left(\frac{\text{nnz}(L)}{\text{TDA} \cdot \text{number of processes}}\right)$ vs. solve time for a variety of large social network graphs on Cori.
    Solves are to a relative tolerance of $10^{-8}$.
    Numeric labels next to points indicate number of processes for a given solve.
    \textit{hollywood} took 15 iterations on 196 processors versus 13 iterations on all other processor sizes, leading to loss of efficiency and negligible speedup.
    Some solves on the same problem perform more iterations (and solve to a slightly higher tolerance) than others causing variation in efficiency.
  }
  \label{fig:efficiency}
\end{figure}
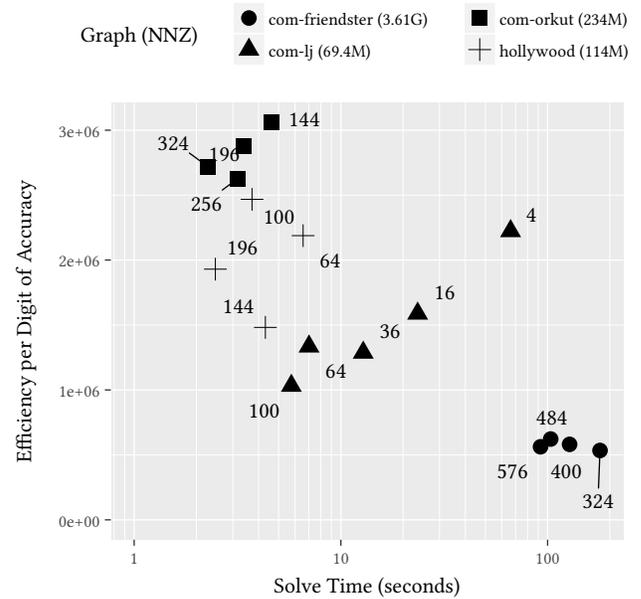

\begin{figure}
  \centering
  \resizebox{\columnwidth}{!}{\input{efficiency_wda}}
  \vspace{-2.5em}
  \caption{
    Semilog-x plot of normalized efficiency $\left(\frac{\text{nnz}(L)}{\text{time per work unit} \cdot \text{number of processes}}\right)$ vs time per digit of accuracy for a variety of large social network graphs on Cori. Numeric labels next to points indicate number of processors used for a given solve.
  }
  \label{fig:efficiency_wda}
\end{figure}
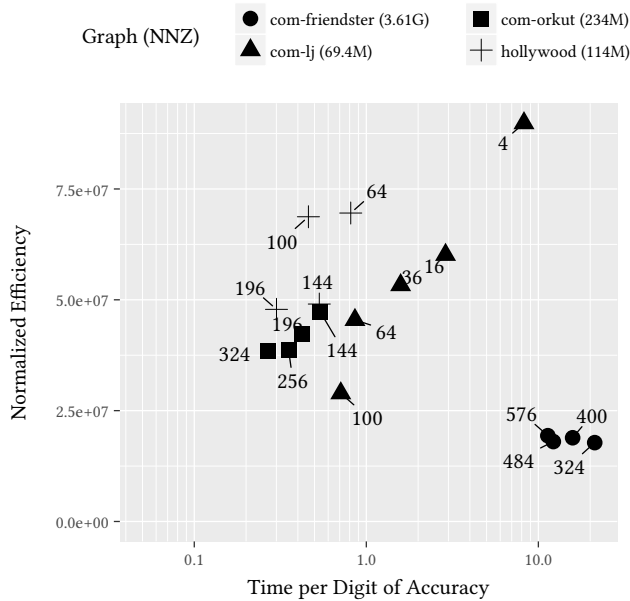

To measure the scalability of our approach, we measured strong scaling on four real world social network graphs using up to 576 processors.
We choose these four graphs because they are some of the largest real world irregular graphs that we can find and are infeasible to solve on a single process.
Many of the graphs in the 110 we use for serial experiments are small enough that solving them on a single process is fast enough.

Our largest graph, \textit{com-friendster}, (from the Stanford Large Network Dataset
Collection \citep{SNAP,friendster}) has 3.6 billion nonzero entries in its
Laplacian matrix. Solving a Laplacian of this size on a single node is
infeasible (even if it could fit in memory, the solve time would be much too
long).
Figure~\ref{fig:efficiency} shows the efficiency, measured as
$$\frac{\text{nnz}(L)}{\text{TDA} \cdot \text{number of processes}} \quad \text{TDA} \coloneqq \frac{-\text{time}}{\log_{10}\Delta r}$$
versus solve time (where TDA is time per digit of accuracy).
We do not use a percentage for efficiency because the choice of base efficiency makes it difficult to compare different solves on different graphs.
A horizontal line indicates that the solver is scaling optimally (increasing machine size moves to the left).
The largest graph (\textit{com-friendster}) appears to have better scaling than the smaller graphs.
\textit{com-orkut} and \textit{hollywood} have a much higher efficiency than \textit{com-friendster} and \textit{com-lj} because \textit{com-orkut} and \textit{hollywood} have a lower WDA.
If we normalize by WDA on both axes, so the y axis becomes
$$\frac{\text{nnz}(L) }{\frac{\text{solve time}}{\text{work unit}} \cdot \text{number of processes}}$$
(this is similar to a ``per iteration'' metric), then we get Figure~\ref{fig:efficiency_wda}.
Normalizing by WDA compares efficiency independent of how difficult it is to solve each graph.
The smaller graphs are solved faster but scale poorly with increasing number of processes.
\textit{com-friendster} takes longer to solve because it is larger and efficiency is somewhat lower (likely due to poor cache behavior with the random ordering) but scales better than the smaller graphs.
The poor scaling for smaller graphs is explained by having less work per process, resulting in a solve phase dominated by communication.

Our high setup cost (relative to solve time) also increases the number of repeated solves necessary.
The majority of the setup phase is spent in the finest level triple product ($P^T L P$), which is handled by a generic CombBLAS matrix-matrix product using $+.*$.
A matrix-matrix product that exploits the special structure of $R$ and $P$ would lower setup times.
Still, these setup times are reasonable and can be amortized over multiple solve phases (when possible).

For a complete view of the scalability of our solver, we would like to measure
weak scaling. However, it is difficult to find a fair way to measure weak scaling on
graphs. We could generate a series of increasingly large random graphs, but in
practice, most solvers perform much better on random graphs than real world
graphs.

\section{Conclusions}\label{sec:conc}

We have presented a distributed memory graph Laplacian solver for social network graphs.
Our solver uses a 2D matrix distribution combined with parallel elimination and unsmoothed aggregation to demonstrate parallel performance on up to 576 processes.
Our novel aggregation algorithm can handle arbitrary matrix distributions and forms accurate aggregates while controlling coarse grid complexity on a variety of irregular graphs.
The parallel elimination algorithm presented uses generalized matrix products to find elimination candidates independent of matrix distribution.
To our knowledge, this is the first distributed memory multigrid solver for graph Laplacians.
It enables solving graph Laplacians that would be infeasible to solve on a single computer.

Our solver's robustness is behind that of serial LAMG (as outlined in section \ref{sec:comparelamg}) but outperforms the natural parallel extensions of LAMG (No iterate recombination, Jacobi smoothing, and cycle index 1).
Further improvement is a topic for future research.
Our solver performs well on social network graphs, but WDA is sometimes high for more regular graphs such as road networks.
Many such graphs also admit a vertex partition with low edge cut, in which case our 2D distribution may be unnecessary.
Extending our solver to handle scaled graph Laplacians (used by some applications) and unsymmetric graphs are other areas for future research.

\begin{acks}

  This material is based upon work supported by the \grantsponsor{0}{U.S. Department of Energy, Office of Science, Office of Advanced Scientific Computing Research}{} under Award Number~\grantnum{0}{DE-SC0016140}.
  This research used resources of the \grantsponsor{1}{National Energy Research Scientific Computing Center, a DOE Office of Science User Facility supported by the Office of Science of the U.S. Department of Energy}{} under Contract No.~\grantnum{1}{DE-AC02-05CH11231}.

\end{acks}

\bibliographystyle{ACM-Reference-Format}
\bibliography{references}

\end{document}

%% file: serial_box_lamg.tex
\begin{tikzpicture}[x=1pt,y=1pt]
\definecolor{fillColor}{RGB}{255,255,255}
\path[use as bounding box,fill=fillColor,fill opacity=0.00] (0,0) rectangle (505.89,144.54);
\begin{scope}
\path[clip] (  0.00,  0.00) rectangle (505.89,144.54);
\definecolor{drawColor}{RGB}{255,255,255}
\definecolor{fillColor}{RGB}{255,255,255}

\path[draw=drawColor,line width= 0.6pt,line join=round,line cap=round,fill=fillColor] (  0.00,  0.00) rectangle (505.89,144.54);
\end{scope}
\begin{scope}
\path[clip] (176.32, 39.22) rectangle (500.39,139.04);
\definecolor{fillColor}{gray}{0.92}

\path[fill=fillColor] (176.32, 39.22) rectangle (500.39,139.04);
\definecolor{drawColor}{RGB}{255,255,255}

\path[draw=drawColor,line width= 0.3pt,line join=round] (227.87, 39.22) --
	(227.87,139.04);

\path[draw=drawColor,line width= 0.3pt,line join=round] (301.53, 39.22) --
	(301.53,139.04);

\path[draw=drawColor,line width= 0.3pt,line join=round] (375.18, 39.22) --
	(375.18,139.04);

\path[draw=drawColor,line width= 0.3pt,line join=round] (448.83, 39.22) --
	(448.83,139.04);

\path[draw=drawColor,line width= 0.6pt,line join=round] (176.32, 46.53) --
	(500.39, 46.53);

\path[draw=drawColor,line width= 0.6pt,line join=round] (176.32, 58.70) --
	(500.39, 58.70);

\path[draw=drawColor,line width= 0.6pt,line join=round] (176.32, 70.87) --
	(500.39, 70.87);

\path[draw=drawColor,line width= 0.6pt,line join=round] (176.32, 83.04) --
	(500.39, 83.04);

\path[draw=drawColor,line width= 0.6pt,line join=round] (176.32, 95.22) --
	(500.39, 95.22);

\path[draw=drawColor,line width= 0.6pt,line join=round] (176.32,107.39) --
	(500.39,107.39);

\path[draw=drawColor,line width= 0.6pt,line join=round] (176.32,119.56) --
	(500.39,119.56);

\path[draw=drawColor,line width= 0.6pt,line join=round] (176.32,131.74) --
	(500.39,131.74);

\path[draw=drawColor,line width= 0.6pt,line join=round] (191.05, 39.22) --
	(191.05,139.04);

\path[draw=drawColor,line width= 0.6pt,line join=round] (264.70, 39.22) --
	(264.70,139.04);

\path[draw=drawColor,line width= 0.6pt,line join=round] (338.35, 39.22) --
	(338.35,139.04);

\path[draw=drawColor,line width= 0.6pt,line join=round] (412.01, 39.22) --
	(412.01,139.04);

\path[draw=drawColor,line width= 0.6pt,line join=round] (485.66, 39.22) --
	(485.66,139.04);
\definecolor{drawColor}{gray}{0.20}
\definecolor{fillColor}{gray}{0.20}

\path[draw=drawColor,line width= 0.4pt,line join=round,line cap=round,fill=fillColor] (290.96, 46.53) circle (  1.96);

\path[draw=drawColor,line width= 0.4pt,line join=round,line cap=round,fill=fillColor] (251.04, 46.53) circle (  1.96);

\path[draw=drawColor,line width= 0.4pt,line join=round,line cap=round,fill=fillColor] (270.09, 46.53) circle (  1.96);

\path[draw=drawColor,line width= 0.4pt,line join=round,line cap=round,fill=fillColor] (255.69, 46.53) circle (  1.96);

\path[draw=drawColor,line width= 0.6pt,line join=round] (222.35, 46.53) -- (242.21, 46.53);

\path[draw=drawColor,line width= 0.6pt,line join=round] (203.46, 46.53) -- (191.05, 46.53);
\definecolor{fillColor}{RGB}{255,255,255}

\path[draw=drawColor,line width= 0.6pt,line join=round,line cap=round,fill=fillColor] (222.35, 41.96) --
	(203.46, 41.96) --
	(203.46, 51.09) --
	(222.35, 51.09) --
	(222.35, 41.96) --
	cycle;

\path[draw=drawColor,line width= 1.1pt,line join=round] (208.78, 41.96) -- (208.78, 51.09);
\definecolor{fillColor}{gray}{0.20}

\path[draw=drawColor,line width= 0.4pt,line join=round,line cap=round,fill=fillColor] (443.07, 58.70) circle (  1.96);

\path[draw=drawColor,line width= 0.6pt,line join=round] (286.52, 58.70) -- (367.11, 58.70);

\path[draw=drawColor,line width= 0.6pt,line join=round] (215.40, 58.70) -- (191.05, 58.70);
\definecolor{fillColor}{RGB}{255,255,255}

\path[draw=drawColor,line width= 0.6pt,line join=round,line cap=round,fill=fillColor] (286.52, 54.13) --
	(215.40, 54.13) --
	(215.40, 63.26) --
	(286.52, 63.26) --
	(286.52, 54.13) --
	cycle;

\path[draw=drawColor,line width= 1.1pt,line join=round] (236.05, 54.13) -- (236.05, 63.26);
\definecolor{fillColor}{gray}{0.20}

\path[draw=drawColor,line width= 0.4pt,line join=round,line cap=round,fill=fillColor] (382.90, 70.87) circle (  1.96);

\path[draw=drawColor,line width= 0.6pt,line join=round] (271.85, 70.87) -- (343.29, 70.87);

\path[draw=drawColor,line width= 0.6pt,line join=round] (209.26, 70.87) -- (191.05, 70.87);
\definecolor{fillColor}{RGB}{255,255,255}

\path[draw=drawColor,line width= 0.6pt,line join=round,line cap=round,fill=fillColor] (271.85, 66.31) --
	(209.26, 66.31) --
	(209.26, 75.44) --
	(271.85, 75.44) --
	(271.85, 66.31) --
	cycle;

\path[draw=drawColor,line width= 1.1pt,line join=round] (221.91, 66.31) -- (221.91, 75.44);

\path[draw=drawColor,line width= 0.6pt,line join=round] (328.57, 83.04) -- (457.83, 83.04);

\path[draw=drawColor,line width= 0.6pt,line join=round] (231.74, 83.04) -- (191.05, 83.04);

\path[draw=drawColor,line width= 0.6pt,line join=round,line cap=round,fill=fillColor] (328.57, 78.48) --
	(231.74, 78.48) --
	(231.74, 87.61) --
	(328.57, 87.61) --
	(328.57, 78.48) --
	cycle;

\path[draw=drawColor,line width= 1.1pt,line join=round] (258.79, 78.48) -- (258.79, 87.61);
\definecolor{fillColor}{gray}{0.20}

\path[draw=drawColor,line width= 0.4pt,line join=round,line cap=round,fill=fillColor] (257.67, 95.22) circle (  1.96);

\path[draw=drawColor,line width= 0.4pt,line join=round,line cap=round,fill=fillColor] (298.47, 95.22) circle (  1.96);

\path[draw=drawColor,line width= 0.6pt,line join=round] (224.06, 95.22) -- (250.69, 95.22);

\path[draw=drawColor,line width= 0.6pt,line join=round] (204.17, 95.22) -- (191.05, 95.22);
\definecolor{fillColor}{RGB}{255,255,255}

\path[draw=drawColor,line width= 0.6pt,line join=round,line cap=round,fill=fillColor] (224.06, 90.65) --
	(204.17, 90.65) --
	(204.17, 99.78) --
	(224.06, 99.78) --
	(224.06, 90.65) --
	cycle;

\path[draw=drawColor,line width= 1.1pt,line join=round] (216.70, 90.65) -- (216.70, 99.78);
\definecolor{fillColor}{gray}{0.20}

\path[draw=drawColor,line width= 0.4pt,line join=round,line cap=round,fill=fillColor] (323.19,107.39) circle (  1.96);

\path[draw=drawColor,line width= 0.4pt,line join=round,line cap=round,fill=fillColor] (308.70,107.39) circle (  1.96);

\path[draw=drawColor,line width= 0.4pt,line join=round,line cap=round,fill=fillColor] (289.96,107.39) circle (  1.96);

\path[draw=drawColor,line width= 0.4pt,line join=round,line cap=round,fill=fillColor] (298.41,107.39) circle (  1.96);

\path[draw=drawColor,line width= 0.4pt,line join=round,line cap=round,fill=fillColor] (443.68,107.39) circle (  1.96);

\path[draw=drawColor,line width= 0.4pt,line join=round,line cap=round,fill=fillColor] (308.26,107.39) circle (  1.96);

\path[draw=drawColor,line width= 0.6pt,line join=round] (246.51,107.39) -- (288.13,107.39);

\path[draw=drawColor,line width= 0.6pt,line join=round] (217.82,107.39) -- (191.05,107.39);
\definecolor{fillColor}{RGB}{255,255,255}

\path[draw=drawColor,line width= 0.6pt,line join=round,line cap=round,fill=fillColor] (246.51,102.83) --
	(217.82,102.83) --
	(217.82,111.96) --
	(246.51,111.96) --
	(246.51,102.83) --
	cycle;

\path[draw=drawColor,line width= 1.1pt,line join=round] (232.49,102.83) -- (232.49,111.96);
\definecolor{fillColor}{gray}{0.20}

\path[draw=drawColor,line width= 0.4pt,line join=round,line cap=round,fill=fillColor] (373.03,119.56) circle (  1.96);

\path[draw=drawColor,line width= 0.6pt,line join=round] (258.21,119.56) -- (312.98,119.56);

\path[draw=drawColor,line width= 0.6pt,line join=round] (210.56,119.56) -- (191.05,119.56);
\definecolor{fillColor}{RGB}{255,255,255}

\path[draw=drawColor,line width= 0.6pt,line join=round,line cap=round,fill=fillColor] (258.21,115.00) --
	(210.56,115.00) --
	(210.56,124.13) --
	(258.21,124.13) --
	(258.21,115.00) --
	cycle;

\path[draw=drawColor,line width= 1.1pt,line join=round] (228.08,115.00) -- (228.08,124.13);
\definecolor{fillColor}{gray}{0.20}

\path[draw=drawColor,line width= 0.4pt,line join=round,line cap=round,fill=fillColor] (398.01,131.74) circle (  1.96);

\path[draw=drawColor,line width= 0.4pt,line join=round,line cap=round,fill=fillColor] (397.19,131.74) circle (  1.96);

\path[draw=drawColor,line width= 0.4pt,line join=round,line cap=round,fill=fillColor] (422.73,131.74) circle (  1.96);

\path[draw=drawColor,line width= 0.4pt,line join=round,line cap=round,fill=fillColor] (458.40,131.74) circle (  1.96);

\path[draw=drawColor,line width= 0.4pt,line join=round,line cap=round,fill=fillColor] (467.10,131.74) circle (  1.96);

\path[draw=drawColor,line width= 0.6pt,line join=round] (300.77,131.74) -- (384.94,131.74);

\path[draw=drawColor,line width= 0.6pt,line join=round] (242.75,131.74) -- (191.05,131.74);
\definecolor{fillColor}{RGB}{255,255,255}

\path[draw=drawColor,line width= 0.6pt,line join=round,line cap=round,fill=fillColor] (300.77,127.17) --
	(242.75,127.17) --
	(242.75,136.30) --
	(300.77,136.30) --
	(300.77,127.17) --
	cycle;

\path[draw=drawColor,line width= 1.1pt,line join=round] (271.07,127.17) -- (271.07,136.30);
\definecolor{drawColor}{RGB}{0,0,0}
\definecolor{fillColor}{RGB}{0,0,0}

\path[draw=drawColor,line width= 0.4pt,line join=round,line cap=round,fill=fillColor] (208.78, 46.53) circle (  2.50);

\path[draw=drawColor,line width= 0.4pt,line join=round,line cap=round,fill=fillColor] (236.05, 58.70) circle (  2.50);

\path[draw=drawColor,line width= 0.4pt,line join=round,line cap=round,fill=fillColor] (221.91, 70.87) circle (  2.50);

\path[draw=drawColor,line width= 0.4pt,line join=round,line cap=round,fill=fillColor] (258.79, 83.04) circle (  2.50);

\path[draw=drawColor,line width= 0.4pt,line join=round,line cap=round,fill=fillColor] (216.70, 95.22) circle (  2.50);

\path[draw=drawColor,line width= 0.4pt,line join=round,line cap=round,fill=fillColor] (232.49,107.39) circle (  2.50);

\path[draw=drawColor,line width= 0.4pt,line join=round,line cap=round,fill=fillColor] (228.08,119.56) circle (  2.50);

\path[draw=drawColor,line width= 0.4pt,line join=round,line cap=round,fill=fillColor] (271.07,131.74) circle (  2.50);
\end{scope}
\begin{scope}
\path[clip] (  0.00,  0.00) rectangle (505.89,144.54);
\definecolor{drawColor}{gray}{0.30}

\node[text=drawColor,anchor=base east,inner sep=0pt, outer sep=0pt, scale=  0.88] at (171.37, 43.50) {Gauss-Seidel.Iterate Recombination.1};

\node[text=drawColor,anchor=base east,inner sep=0pt, outer sep=0pt, scale=  0.88] at (171.37, 55.67) {Jacobi.Iterate Recombination.1};

\node[text=drawColor,anchor=base east,inner sep=0pt, outer sep=0pt, scale=  0.88] at (171.37, 67.84) {Gauss-Seidel.No Recombination.1};

\node[text=drawColor,anchor=base east,inner sep=0pt, outer sep=0pt, scale=  0.88] at (171.37, 80.01) {Jacobi.No Recombination.1};

\node[text=drawColor,anchor=base east,inner sep=0pt, outer sep=0pt, scale=  0.88] at (171.37, 92.19) {Gauss-Seidel.Iterate Recombination.1.5};

\node[text=drawColor,anchor=base east,inner sep=0pt, outer sep=0pt, scale=  0.88] at (171.37,104.36) {Jacobi.Iterate Recombination.1.5};

\node[text=drawColor,anchor=base east,inner sep=0pt, outer sep=0pt, scale=  0.88] at (171.37,116.53) {Gauss-Seidel.No Recombination.1.5};

\node[text=drawColor,anchor=base east,inner sep=0pt, outer sep=0pt, scale=  0.88] at (171.37,128.71) {Jacobi.No Recombination.1.5};
\end{scope}
\begin{scope}
\path[clip] (  0.00,  0.00) rectangle (505.89,144.54);
\definecolor{drawColor}{gray}{0.20}

\path[draw=drawColor,line width= 0.6pt,line join=round] (173.57, 46.53) --
	(176.32, 46.53);

\path[draw=drawColor,line width= 0.6pt,line join=round] (173.57, 58.70) --
	(176.32, 58.70);

\path[draw=drawColor,line width= 0.6pt,line join=round] (173.57, 70.87) --
	(176.32, 70.87);

\path[draw=drawColor,line width= 0.6pt,line join=round] (173.57, 83.04) --
	(176.32, 83.04);

\path[draw=drawColor,line width= 0.6pt,line join=round] (173.57, 95.22) --
	(176.32, 95.22);

\path[draw=drawColor,line width= 0.6pt,line join=round] (173.57,107.39) --
	(176.32,107.39);

\path[draw=drawColor,line width= 0.6pt,line join=round] (173.57,119.56) --
	(176.32,119.56);

\path[draw=drawColor,line width= 0.6pt,line join=round] (173.57,131.74) --
	(176.32,131.74);
\end{scope}
\begin{scope}
\path[clip] (  0.00,  0.00) rectangle (505.89,144.54);
\definecolor{drawColor}{gray}{0.20}

\path[draw=drawColor,line width= 0.6pt,line join=round] (191.05, 36.47) --
	(191.05, 39.22);

\path[draw=drawColor,line width= 0.6pt,line join=round] (264.70, 36.47) --
	(264.70, 39.22);

\path[draw=drawColor,line width= 0.6pt,line join=round] (338.35, 36.47) --
	(338.35, 39.22);

\path[draw=drawColor,line width= 0.6pt,line join=round] (412.01, 36.47) --
	(412.01, 39.22);

\path[draw=drawColor,line width= 0.6pt,line join=round] (485.66, 36.47) --
	(485.66, 39.22);
\end{scope}
\begin{scope}
\path[clip] (  0.00,  0.00) rectangle (505.89,144.54);
\definecolor{drawColor}{gray}{0.30}

\node[text=drawColor,rotate= 90.00,anchor=base east,inner sep=0pt, outer sep=0pt, scale=  0.88] at (197.11, 34.27) {0};

\node[text=drawColor,rotate= 90.00,anchor=base east,inner sep=0pt, outer sep=0pt, scale=  0.88] at (270.76, 34.27) {25};

\node[text=drawColor,rotate= 90.00,anchor=base east,inner sep=0pt, outer sep=0pt, scale=  0.88] at (344.41, 34.27) {50};

\node[text=drawColor,rotate= 90.00,anchor=base east,inner sep=0pt, outer sep=0pt, scale=  0.88] at (418.07, 34.27) {75};

\node[text=drawColor,rotate= 90.00,anchor=base east,inner sep=0pt, outer sep=0pt, scale=  0.88] at (491.72, 34.27) {100};
\end{scope}
\begin{scope}
\path[clip] (  0.00,  0.00) rectangle (505.89,144.54);
\definecolor{drawColor}{RGB}{0,0,0}

\node[text=drawColor,anchor=base,inner sep=0pt, outer sep=0pt, scale=  1.10] at (338.35,  8.00) {WDA (Smaller is better)};
\end{scope}
\begin{scope}
\path[clip] (  0.00,  0.00) rectangle (505.89,144.54);
\definecolor{drawColor}{RGB}{0,0,0}

\node[text=drawColor,rotate= 90.00,anchor=base,inner sep=0pt, outer sep=0pt, scale=  1.10] at ( 13.08, 89.13) {Solver Configuration};
\end{scope}
\end{tikzpicture}

%% file: serial_box.tex
\begin{tikzpicture}[x=1pt,y=1pt]
\definecolor{fillColor}{RGB}{255,255,255}
\path[use as bounding box,fill=fillColor,fill opacity=0.00] (0,0) rectangle (578.16,130.09);
\begin{scope}
\path[clip] (  0.00,  0.00) rectangle (578.16,130.09);
\definecolor{drawColor}{RGB}{255,255,255}
\definecolor{fillColor}{RGB}{255,255,255}

\path[draw=drawColor,line width= 0.6pt,line join=round,line cap=round,fill=fillColor] (  0.00,  0.00) rectangle (578.16,130.09);
\end{scope}
\begin{scope}
\path[clip] (205.83, 39.22) rectangle (572.66,124.59);
\definecolor{fillColor}{gray}{0.92}

\path[fill=fillColor] (205.83, 39.22) rectangle (572.66,124.59);
\definecolor{drawColor}{RGB}{255,255,255}

\path[draw=drawColor,line width= 0.3pt,line join=round] (264.19, 39.22) --
	(264.19,124.59);

\path[draw=drawColor,line width= 0.3pt,line join=round] (347.56, 39.22) --
	(347.56,124.59);

\path[draw=drawColor,line width= 0.3pt,line join=round] (430.93, 39.22) --
	(430.93,124.59);

\path[draw=drawColor,line width= 0.3pt,line join=round] (514.30, 39.22) --
	(514.30,124.59);

\path[draw=drawColor,line width= 0.6pt,line join=round] (205.83, 51.42) --
	(572.66, 51.42);

\path[draw=drawColor,line width= 0.6pt,line join=round] (205.83, 71.74) --
	(572.66, 71.74);

\path[draw=drawColor,line width= 0.6pt,line join=round] (205.83, 92.07) --
	(572.66, 92.07);

\path[draw=drawColor,line width= 0.6pt,line join=round] (205.83,112.39) --
	(572.66,112.39);

\path[draw=drawColor,line width= 0.6pt,line join=round] (222.50, 39.22) --
	(222.50,124.59);

\path[draw=drawColor,line width= 0.6pt,line join=round] (305.87, 39.22) --
	(305.87,124.59);

\path[draw=drawColor,line width= 0.6pt,line join=round] (389.24, 39.22) --
	(389.24,124.59);

\path[draw=drawColor,line width= 0.6pt,line join=round] (472.61, 39.22) --
	(472.61,124.59);

\path[draw=drawColor,line width= 0.6pt,line join=round] (555.99, 39.22) --
	(555.99,124.59);
\definecolor{drawColor}{gray}{0.20}
\definecolor{fillColor}{gray}{0.20}

\path[draw=drawColor,line width= 0.4pt,line join=round,line cap=round,fill=fillColor] (297.91, 51.42) circle (  1.96);

\path[draw=drawColor,line width= 0.4pt,line join=round,line cap=round,fill=fillColor] (344.10, 51.42) circle (  1.96);

\path[draw=drawColor,line width= 0.6pt,line join=round] (259.87, 51.42) -- (290.01, 51.42);

\path[draw=drawColor,line width= 0.6pt,line join=round] (237.35, 51.42) -- (222.50, 51.42);
\definecolor{fillColor}{RGB}{255,255,255}

\path[draw=drawColor,line width= 0.6pt,line join=round,line cap=round,fill=fillColor] (259.87, 43.80) --
	(237.35, 43.80) --
	(237.35, 59.04) --
	(259.87, 59.04) --
	(259.87, 43.80) --
	cycle;

\path[draw=drawColor,line width= 1.1pt,line join=round] (251.54, 43.80) -- (251.54, 59.04);

\path[draw=drawColor,line width= 0.6pt,line join=round] (378.17, 71.74) -- (524.48, 71.74);

\path[draw=drawColor,line width= 0.6pt,line join=round] (268.57, 71.74) -- (222.50, 71.74);

\path[draw=drawColor,line width= 0.6pt,line join=round,line cap=round,fill=fillColor] (378.17, 64.12) --
	(268.57, 64.12) --
	(268.57, 79.36) --
	(378.17, 79.36) --
	(378.17, 64.12) --
	cycle;

\path[draw=drawColor,line width= 1.1pt,line join=round] (299.19, 64.12) -- (299.19, 79.36);
\definecolor{fillColor}{gray}{0.20}

\path[draw=drawColor,line width= 0.4pt,line join=round,line cap=round,fill=fillColor] (398.58, 92.07) circle (  1.96);

\path[draw=drawColor,line width= 0.4pt,line join=round,line cap=round,fill=fillColor] (398.01, 92.07) circle (  1.96);

\path[draw=drawColor,line width= 0.4pt,line join=round,line cap=round,fill=fillColor] (395.86, 92.07) circle (  1.96);

\path[draw=drawColor,line width= 0.4pt,line join=round,line cap=round,fill=fillColor] (408.12, 92.07) circle (  1.96);

\path[draw=drawColor,line width= 0.4pt,line join=round,line cap=round,fill=fillColor] (406.61, 92.07) circle (  1.96);

\path[draw=drawColor,line width= 0.4pt,line join=round,line cap=round,fill=fillColor] (430.36, 92.07) circle (  1.96);

\path[draw=drawColor,line width= 0.4pt,line join=round,line cap=round,fill=fillColor] (428.56, 92.07) circle (  1.96);

\path[draw=drawColor,line width= 0.4pt,line join=round,line cap=round,fill=fillColor] (392.86, 92.07) circle (  1.96);

\path[draw=drawColor,line width= 0.4pt,line join=round,line cap=round,fill=fillColor] (393.28, 92.07) circle (  1.96);

\path[draw=drawColor,line width= 0.4pt,line join=round,line cap=round,fill=fillColor] (393.81, 92.07) circle (  1.96);

\path[draw=drawColor,line width= 0.4pt,line join=round,line cap=round,fill=fillColor] (439.32, 92.07) circle (  1.96);

\path[draw=drawColor,line width= 0.4pt,line join=round,line cap=round,fill=fillColor] (440.67, 92.07) circle (  1.96);

\path[draw=drawColor,line width= 0.4pt,line join=round,line cap=round,fill=fillColor] (436.58, 92.07) circle (  1.96);

\path[draw=drawColor,line width= 0.4pt,line join=round,line cap=round,fill=fillColor] (397.05, 92.07) circle (  1.96);

\path[draw=drawColor,line width= 0.6pt,line join=round] (314.64, 92.07) -- (381.56, 92.07);

\path[draw=drawColor,line width= 0.6pt,line join=round] (262.73, 92.07) -- (232.77, 92.07);
\definecolor{fillColor}{RGB}{255,255,255}

\path[draw=drawColor,line width= 0.6pt,line join=round,line cap=round,fill=fillColor] (314.64, 84.44) --
	(262.73, 84.44) --
	(262.73, 99.69) --
	(314.64, 99.69) --
	(314.64, 84.44) --
	cycle;

\path[draw=drawColor,line width= 1.1pt,line join=round] (275.63, 84.44) -- (275.63, 99.69);
\definecolor{fillColor}{gray}{0.20}

\path[draw=drawColor,line width= 0.4pt,line join=round,line cap=round,fill=fillColor] (451.37,112.39) circle (  1.96);

\path[draw=drawColor,line width= 0.4pt,line join=round,line cap=round,fill=fillColor] (457.72,112.39) circle (  1.96);

\path[draw=drawColor,line width= 0.4pt,line join=round,line cap=round,fill=fillColor] (456.25,112.39) circle (  1.96);

\path[draw=drawColor,line width= 0.4pt,line join=round,line cap=round,fill=fillColor] (533.86,112.39) circle (  1.96);

\path[draw=drawColor,line width= 0.4pt,line join=round,line cap=round,fill=fillColor] (532.27,112.39) circle (  1.96);

\path[draw=drawColor,line width= 0.4pt,line join=round,line cap=round,fill=fillColor] (528.40,112.39) circle (  1.96);

\path[draw=drawColor,line width= 0.4pt,line join=round,line cap=round,fill=fillColor] (446.58,112.39) circle (  1.96);

\path[draw=drawColor,line width= 0.4pt,line join=round,line cap=round,fill=fillColor] (449.23,112.39) circle (  1.96);

\path[draw=drawColor,line width= 0.4pt,line join=round,line cap=round,fill=fillColor] (553.32,112.39) circle (  1.96);

\path[draw=drawColor,line width= 0.4pt,line join=round,line cap=round,fill=fillColor] (555.21,112.39) circle (  1.96);

\path[draw=drawColor,line width= 0.4pt,line join=round,line cap=round,fill=fillColor] (488.78,112.39) circle (  1.96);

\path[draw=drawColor,line width= 0.4pt,line join=round,line cap=round,fill=fillColor] (453.40,112.39) circle (  1.96);

\path[draw=drawColor,line width= 0.4pt,line join=round,line cap=round,fill=fillColor] (464.20,112.39) circle (  1.96);

\path[draw=drawColor,line width= 0.4pt,line join=round,line cap=round,fill=fillColor] (541.02,112.39) circle (  1.96);

\path[draw=drawColor,line width= 0.4pt,line join=round,line cap=round,fill=fillColor] (538.26,112.39) circle (  1.96);

\path[draw=drawColor,line width= 0.4pt,line join=round,line cap=round,fill=fillColor] (546.34,112.39) circle (  1.96);

\path[draw=drawColor,line width= 0.4pt,line join=round,line cap=round,fill=fillColor] (475.74,112.39) circle (  1.96);

\path[draw=drawColor,line width= 0.4pt,line join=round,line cap=round,fill=fillColor] (481.94,112.39) circle (  1.96);

\path[draw=drawColor,line width= 0.4pt,line join=round,line cap=round,fill=fillColor] (477.54,112.39) circle (  1.96);

\path[draw=drawColor,line width= 0.4pt,line join=round,line cap=round,fill=fillColor] (487.26,112.39) circle (  1.96);

\path[draw=drawColor,line width= 0.4pt,line join=round,line cap=round,fill=fillColor] (494.45,112.39) circle (  1.96);

\path[draw=drawColor,line width= 0.4pt,line join=round,line cap=round,fill=fillColor] (486.12,112.39) circle (  1.96);

\path[draw=drawColor,line width= 0.4pt,line join=round,line cap=round,fill=fillColor] (539.32,112.39) circle (  1.96);

\path[draw=drawColor,line width= 0.4pt,line join=round,line cap=round,fill=fillColor] (541.14,112.39) circle (  1.96);

\path[draw=drawColor,line width= 0.4pt,line join=round,line cap=round,fill=fillColor] (535.51,112.39) circle (  1.96);

\path[draw=drawColor,line width= 0.4pt,line join=round,line cap=round,fill=fillColor] (472.47,112.39) circle (  1.96);

\path[draw=drawColor,line width= 0.4pt,line join=round,line cap=round,fill=fillColor] (481.36,112.39) circle (  1.96);

\path[draw=drawColor,line width= 0.4pt,line join=round,line cap=round,fill=fillColor] (469.18,112.39) circle (  1.96);

\path[draw=drawColor,line width= 0.6pt,line join=round] (325.09,112.39) -- (415.17,112.39);

\path[draw=drawColor,line width= 0.6pt,line join=round] (247.95,112.39) -- (230.01,112.39);
\definecolor{fillColor}{RGB}{255,255,255}

\path[draw=drawColor,line width= 0.6pt,line join=round,line cap=round,fill=fillColor] (325.09,104.77) --
	(247.95,104.77) --
	(247.95,120.01) --
	(325.09,120.01) --
	(325.09,104.77) --
	cycle;

\path[draw=drawColor,line width= 1.1pt,line join=round] (279.19,104.77) -- (279.19,120.01);
\definecolor{drawColor}{RGB}{0,0,0}
\definecolor{fillColor}{RGB}{0,0,0}

\path[draw=drawColor,line width= 0.4pt,line join=round,line cap=round,fill=fillColor] (251.54, 51.42) circle (  1.96);

\path[draw=drawColor,line width= 0.4pt,line join=round,line cap=round,fill=fillColor] (299.19, 71.74) circle (  1.96);

\path[draw=drawColor,line width= 0.4pt,line join=round,line cap=round,fill=fillColor] (275.63, 92.07) circle (  1.96);

\path[draw=drawColor,line width= 0.4pt,line join=round,line cap=round,fill=fillColor] (279.19,112.39) circle (  1.96);
\end{scope}
\begin{scope}
\path[clip] (  0.00,  0.00) rectangle (578.16,130.09);
\definecolor{drawColor}{gray}{0.30}

\node[text=drawColor,anchor=base east,inner sep=0pt, outer sep=0pt, scale=  0.88] at (200.88, 48.39) {LAMG.Gauss-Seidel.Iterate Recombination.1.5};

\node[text=drawColor,anchor=base east,inner sep=0pt, outer sep=0pt, scale=  0.88] at (200.88, 68.71) {LAMG.Jacobi.No Recombination.1};

\node[text=drawColor,anchor=base east,inner sep=0pt, outer sep=0pt, scale=  0.88] at (200.88, 89.04) {Our Solver.Chebyshev.K-Cycle.1};

\node[text=drawColor,anchor=base east,inner sep=0pt, outer sep=0pt, scale=  0.88] at (200.88,109.36) {Our Solver.Chebyshev.V-Cycle.1};
\end{scope}
\begin{scope}
\path[clip] (  0.00,  0.00) rectangle (578.16,130.09);
\definecolor{drawColor}{gray}{0.20}

\path[draw=drawColor,line width= 0.6pt,line join=round] (203.08, 51.42) --
	(205.83, 51.42);

\path[draw=drawColor,line width= 0.6pt,line join=round] (203.08, 71.74) --
	(205.83, 71.74);

\path[draw=drawColor,line width= 0.6pt,line join=round] (203.08, 92.07) --
	(205.83, 92.07);

\path[draw=drawColor,line width= 0.6pt,line join=round] (203.08,112.39) --
	(205.83,112.39);
\end{scope}
\begin{scope}
\path[clip] (  0.00,  0.00) rectangle (578.16,130.09);
\definecolor{drawColor}{gray}{0.20}

\path[draw=drawColor,line width= 0.6pt,line join=round] (222.50, 36.47) --
	(222.50, 39.22);

\path[draw=drawColor,line width= 0.6pt,line join=round] (305.87, 36.47) --
	(305.87, 39.22);

\path[draw=drawColor,line width= 0.6pt,line join=round] (389.24, 36.47) --
	(389.24, 39.22);

\path[draw=drawColor,line width= 0.6pt,line join=round] (472.61, 36.47) --
	(472.61, 39.22);

\path[draw=drawColor,line width= 0.6pt,line join=round] (555.99, 36.47) --
	(555.99, 39.22);
\end{scope}
\begin{scope}
\path[clip] (  0.00,  0.00) rectangle (578.16,130.09);
\definecolor{drawColor}{gray}{0.30}

\node[text=drawColor,rotate= 90.00,anchor=base east,inner sep=0pt, outer sep=0pt, scale=  0.88] at (228.56, 34.27) {0};

\node[text=drawColor,rotate= 90.00,anchor=base east,inner sep=0pt, outer sep=0pt, scale=  0.88] at (311.93, 34.27) {25};

\node[text=drawColor,rotate= 90.00,anchor=base east,inner sep=0pt, outer sep=0pt, scale=  0.88] at (395.30, 34.27) {50};

\node[text=drawColor,rotate= 90.00,anchor=base east,inner sep=0pt, outer sep=0pt, scale=  0.88] at (478.68, 34.27) {75};

\node[text=drawColor,rotate= 90.00,anchor=base east,inner sep=0pt, outer sep=0pt, scale=  0.88] at (562.05, 34.27) {100};
\end{scope}
\begin{scope}
\path[clip] (  0.00,  0.00) rectangle (578.16,130.09);
\definecolor{drawColor}{RGB}{0,0,0}

\node[text=drawColor,anchor=base,inner sep=0pt, outer sep=0pt, scale=  1.10] at (389.24,  8.00) {WDA (Smaller is better)};
\end{scope}
\begin{scope}
\path[clip] (  0.00,  0.00) rectangle (578.16,130.09);
\definecolor{drawColor}{RGB}{0,0,0}

\node[text=drawColor,rotate= 90.00,anchor=base,inner sep=0pt, outer sep=0pt, scale=  1.10] at ( 13.08, 81.90) {Solver Configuration};
\end{scope}
\end{tikzpicture}

%% file: kcycles.tex
\begin{tikzpicture}[x=1pt,y=1pt]
\definecolor{fillColor}{RGB}{255,255,255}
\path[use as bounding box,fill=fillColor,fill opacity=0.00] (0,0) rectangle (289.08,289.08);
\begin{scope}
\path[clip] (  0.00,  0.00) rectangle (289.08,289.08);
\definecolor{drawColor}{RGB}{255,255,255}
\definecolor{fillColor}{RGB}{255,255,255}

\path[draw=drawColor,line width= 0.6pt,line join=round,line cap=round,fill=fillColor] (  0.00, -0.00) rectangle (289.08,289.08);
\end{scope}
\begin{scope}
\path[clip] ( 39.22, 29.59) rectangle (283.58,246.36);
\definecolor{fillColor}{gray}{0.92}

\path[fill=fillColor] ( 39.22, 29.59) rectangle (283.58,246.36);
\definecolor{drawColor}{RGB}{255,255,255}

\path[draw=drawColor,line width= 0.3pt,line join=round] ( 39.22, 30.70) --
	(283.58, 30.70);

\path[draw=drawColor,line width= 0.3pt,line join=round] ( 39.22, 36.24) --
	(283.58, 36.24);

\path[draw=drawColor,line width= 0.3pt,line join=round] ( 39.22, 41.03) --
	(283.58, 41.03);

\path[draw=drawColor,line width= 0.3pt,line join=round] ( 39.22, 45.26) --
	(283.58, 45.26);

\path[draw=drawColor,line width= 0.3pt,line join=round] ( 39.22, 73.93) --
	(283.58, 73.93);

\path[draw=drawColor,line width= 0.3pt,line join=round] ( 39.22, 88.48) --
	(283.58, 88.48);

\path[draw=drawColor,line width= 0.3pt,line join=round] ( 39.22, 98.81) --
	(283.58, 98.81);

\path[draw=drawColor,line width= 0.3pt,line join=round] ( 39.22,106.82) --
	(283.58,106.82);

\path[draw=drawColor,line width= 0.3pt,line join=round] ( 39.22,113.37) --
	(283.58,113.37);

\path[draw=drawColor,line width= 0.3pt,line join=round] ( 39.22,118.90) --
	(283.58,118.90);

\path[draw=drawColor,line width= 0.3pt,line join=round] ( 39.22,123.69) --
	(283.58,123.69);

\path[draw=drawColor,line width= 0.3pt,line join=round] ( 39.22,127.92) --
	(283.58,127.92);

\path[draw=drawColor,line width= 0.3pt,line join=round] ( 39.22,156.59) --
	(283.58,156.59);

\path[draw=drawColor,line width= 0.3pt,line join=round] ( 39.22,171.14) --
	(283.58,171.14);

\path[draw=drawColor,line width= 0.3pt,line join=round] ( 39.22,181.47) --
	(283.58,181.47);

\path[draw=drawColor,line width= 0.3pt,line join=round] ( 39.22,189.48) --
	(283.58,189.48);

\path[draw=drawColor,line width= 0.3pt,line join=round] ( 39.22,196.03) --
	(283.58,196.03);

\path[draw=drawColor,line width= 0.3pt,line join=round] ( 39.22,201.56) --
	(283.58,201.56);

\path[draw=drawColor,line width= 0.3pt,line join=round] ( 39.22,206.35) --
	(283.58,206.35);

\path[draw=drawColor,line width= 0.3pt,line join=round] ( 39.22,210.58) --
	(283.58,210.58);

\path[draw=drawColor,line width= 0.3pt,line join=round] ( 39.22,239.25) --
	(283.58,239.25);

\path[draw=drawColor,line width= 0.3pt,line join=round] ( 50.33, 29.59) --
	( 50.33,246.36);

\path[draw=drawColor,line width= 0.3pt,line join=round] ( 63.07, 29.59) --
	( 63.07,246.36);

\path[draw=drawColor,line width= 0.3pt,line join=round] ( 73.47, 29.59) --
	( 73.47,246.36);

\path[draw=drawColor,line width= 0.3pt,line join=round] ( 82.27, 29.59) --
	( 82.27,246.36);

\path[draw=drawColor,line width= 0.3pt,line join=round] ( 89.89, 29.59) --
	( 89.89,246.36);

\path[draw=drawColor,line width= 0.3pt,line join=round] ( 96.62, 29.59) --
	( 96.62,246.36);

\path[draw=drawColor,line width= 0.3pt,line join=round] (142.20, 29.59) --
	(142.20,246.36);

\path[draw=drawColor,line width= 0.3pt,line join=round] (165.34, 29.59) --
	(165.34,246.36);

\path[draw=drawColor,line width= 0.3pt,line join=round] (181.76, 29.59) --
	(181.76,246.36);

\path[draw=drawColor,line width= 0.3pt,line join=round] (194.50, 29.59) --
	(194.50,246.36);

\path[draw=drawColor,line width= 0.3pt,line join=round] (204.90, 29.59) --
	(204.90,246.36);

\path[draw=drawColor,line width= 0.3pt,line join=round] (213.70, 29.59) --
	(213.70,246.36);

\path[draw=drawColor,line width= 0.3pt,line join=round] (221.32, 29.59) --
	(221.32,246.36);

\path[draw=drawColor,line width= 0.3pt,line join=round] (228.05, 29.59) --
	(228.05,246.36);

\path[draw=drawColor,line width= 0.3pt,line join=round] (273.63, 29.59) --
	(273.63,246.36);

\path[draw=drawColor,line width= 0.6pt,line join=round] ( 39.22, 49.04) --
	(283.58, 49.04);

\path[draw=drawColor,line width= 0.6pt,line join=round] ( 39.22,131.70) --
	(283.58,131.70);

\path[draw=drawColor,line width= 0.6pt,line join=round] ( 39.22,214.36) --
	(283.58,214.36);

\path[draw=drawColor,line width= 0.6pt,line join=round] (102.63, 29.59) --
	(102.63,246.36);

\path[draw=drawColor,line width= 0.6pt,line join=round] (234.06, 29.59) --
	(234.06,246.36);
\definecolor{fillColor}{RGB}{0,0,0}

\path[fill=fillColor] (208.59,184.57) --
	(213.39,176.24) --
	(203.78,176.24) --
	cycle;

\path[fill=fillColor] (234.06,174.76) --
	(238.87,166.44) --
	(229.26,166.44) --
	cycle;

\path[fill=fillColor] (254.88,146.56) --
	(259.68,138.24) --
	(250.07,138.24) --
	cycle;

\path[fill=fillColor] (272.47,193.30) --
	(277.28,184.98) --
	(267.67,184.98) --
	cycle;

\path[fill=fillColor] (208.59, 84.20) circle (  3.57);

\path[fill=fillColor] (234.06, 70.26) circle (  3.57);

\path[fill=fillColor] (254.88, 59.55) circle (  3.57);

\path[fill=fillColor] (272.47, 59.99) circle (  3.57);

\path[fill=fillColor] ( 50.33,216.77) --
	( 55.14,208.44) --
	( 45.52,208.44) --
	cycle;

\path[fill=fillColor] (129.46,242.06) --
	(134.26,233.74) --
	(124.65,233.74) --
	cycle;

\path[fill=fillColor] (175.75,190.22) --
	(180.55,181.90) --
	(170.94,181.90) --
	cycle;

\path[fill=fillColor] ( 50.33,179.41) circle (  3.57);

\path[fill=fillColor] (129.46,125.70) circle (  3.57);

\path[fill=fillColor] (175.75,110.10) circle (  3.57);
\definecolor{drawColor}{RGB}{0,0,0}

\node[text=drawColor,anchor=base,inner sep=0pt, outer sep=0pt, scale=  1.14] at (201.51,167.97) {64};

\node[text=drawColor,anchor=base,inner sep=0pt, outer sep=0pt, scale=  1.14] at (225.43,157.88) {100};

\node[text=drawColor,anchor=base,inner sep=0pt, outer sep=0pt, scale=  1.14] at (247.23,129.95) {144};

\node[text=drawColor,anchor=base,inner sep=0pt, outer sep=0pt, scale=  1.14] at (260.75,183.89) {196};

\node[text=drawColor,anchor=base,inner sep=0pt, outer sep=0pt, scale=  1.14] at (201.26, 87.12) {64};

\node[text=drawColor,anchor=base,inner sep=0pt, outer sep=0pt, scale=  1.14] at (225.98, 73.28) {100};

\node[text=drawColor,anchor=base,inner sep=0pt, outer sep=0pt, scale=  1.14] at (246.35, 47.81) {144};

\node[text=drawColor,anchor=base,inner sep=0pt, outer sep=0pt, scale=  1.14] at (265.77, 63.08) {196};

\node[text=drawColor,anchor=base,inner sep=0pt, outer sep=0pt, scale=  1.14] at ( 45.10,200.32) {4};

\node[text=drawColor,anchor=base,inner sep=0pt, outer sep=0pt, scale=  1.14] at (138.78,235.50) {16};

\node[text=drawColor,anchor=base,inner sep=0pt, outer sep=0pt, scale=  1.14] at (168.31,173.90) {36};

\node[text=drawColor,anchor=base,inner sep=0pt, outer sep=0pt, scale=  1.14] at ( 57.22,182.23) {4};

\node[text=drawColor,anchor=base,inner sep=0pt, outer sep=0pt, scale=  1.14] at (120.88,114.30) {16};

\node[text=drawColor,anchor=base,inner sep=0pt, outer sep=0pt, scale=  1.14] at (184.14,113.13) {36};

\path[draw=drawColor,line width= 0.6pt,line join=round] ( 50.33,179.15) --
	( 50.33,179.15) --
	(129.46,129.39) --
	(129.46,129.39) --
	(175.75,100.27) --
	(175.75,100.27) --
	(208.59, 79.62) --
	(208.59, 79.62) --
	(234.06, 63.60) --
	(234.06, 63.60) --
	(254.88, 50.51) --
	(254.88, 50.51) --
	(272.47, 39.44) --
	(272.47, 39.44);
\end{scope}
\begin{scope}
\path[clip] (  0.00,  0.00) rectangle (289.08,289.08);
\definecolor{drawColor}{gray}{0.30}

\node[text=drawColor,anchor=base east,inner sep=0pt, outer sep=0pt, scale=  0.88] at ( 34.27, 46.01) {1};

\node[text=drawColor,anchor=base east,inner sep=0pt, outer sep=0pt, scale=  0.88] at ( 34.27,128.67) {10};

\node[text=drawColor,anchor=base east,inner sep=0pt, outer sep=0pt, scale=  0.88] at ( 34.27,211.33) {100};
\end{scope}
\begin{scope}
\path[clip] (  0.00,  0.00) rectangle (289.08,289.08);
\definecolor{drawColor}{gray}{0.20}

\path[draw=drawColor,line width= 0.6pt,line join=round] ( 36.47, 49.04) --
	( 39.22, 49.04);

\path[draw=drawColor,line width= 0.6pt,line join=round] ( 36.47,131.70) --
	( 39.22,131.70);

\path[draw=drawColor,line width= 0.6pt,line join=round] ( 36.47,214.36) --
	( 39.22,214.36);
\end{scope}
\begin{scope}
\path[clip] (  0.00,  0.00) rectangle (289.08,289.08);
\definecolor{drawColor}{gray}{0.20}

\path[draw=drawColor,line width= 0.6pt,line join=round] (102.63, 26.84) --
	(102.63, 29.59);

\path[draw=drawColor,line width= 0.6pt,line join=round] (234.06, 26.84) --
	(234.06, 29.59);
\end{scope}
\begin{scope}
\path[clip] (  0.00,  0.00) rectangle (289.08,289.08);
\definecolor{drawColor}{gray}{0.30}

\node[text=drawColor,anchor=base,inner sep=0pt, outer sep=0pt, scale=  0.88] at (102.63, 18.58) {10};

\node[text=drawColor,anchor=base,inner sep=0pt, outer sep=0pt, scale=  0.88] at (234.06, 18.58) {100};
\end{scope}
\begin{scope}
\path[clip] (  0.00,  0.00) rectangle (289.08,289.08);
\definecolor{drawColor}{RGB}{0,0,0}

\node[text=drawColor,anchor=base,inner sep=0pt, outer sep=0pt, scale=  1.10] at (161.40,  5.50) {Processes};
\end{scope}
\begin{scope}
\path[clip] (  0.00,  0.00) rectangle (289.08,289.08);
\definecolor{drawColor}{RGB}{0,0,0}

\node[text=drawColor,rotate= 90.00,anchor=base,inner sep=0pt, outer sep=0pt, scale=  1.10] at ( 13.08,137.98) {Solve Time (seconds)};
\end{scope}
\begin{scope}
\path[clip] (  0.00,  0.00) rectangle (289.08,289.08);
\definecolor{fillColor}{RGB}{255,255,255}

\path[fill=fillColor] ( 56.48,257.74) rectangle (174.57,283.58);
\end{scope}
\begin{scope}
\path[clip] (  0.00,  0.00) rectangle (289.08,289.08);
\definecolor{drawColor}{RGB}{255,255,255}
\definecolor{fillColor}{gray}{0.95}

\path[draw=drawColor,line width= 0.6pt,line join=round,line cap=round,fill=fillColor] ( 65.78,263.44) rectangle ( 80.23,277.89);
\end{scope}
\begin{scope}
\path[clip] (  0.00,  0.00) rectangle (289.08,289.08);
\definecolor{fillColor}{RGB}{0,0,0}

\path[fill=fillColor] ( 73.01,270.66) circle (  3.57);
\end{scope}
\begin{scope}
\path[clip] (  0.00,  0.00) rectangle (289.08,289.08);
\definecolor{drawColor}{RGB}{255,255,255}
\definecolor{fillColor}{gray}{0.95}

\path[draw=drawColor,line width= 0.6pt,line join=round,line cap=round,fill=fillColor] (118.11,263.44) rectangle (132.56,277.89);
\end{scope}
\begin{scope}
\path[clip] (  0.00,  0.00) rectangle (289.08,289.08);
\definecolor{fillColor}{RGB}{0,0,0}

\path[fill=fillColor] (125.34,276.21) --
	(130.14,267.89) --
	(120.53,267.89) --
	cycle;
\end{scope}
\begin{scope}
\path[clip] (  0.00,  0.00) rectangle (289.08,289.08);
\definecolor{drawColor}{RGB}{0,0,0}

\node[text=drawColor,anchor=base west,inner sep=0pt, outer sep=0pt, scale=  0.88] at ( 82.04,267.63) {V-Cycles};
\end{scope}
\begin{scope}
\path[clip] (  0.00,  0.00) rectangle (289.08,289.08);
\definecolor{drawColor}{RGB}{0,0,0}

\node[text=drawColor,anchor=base west,inner sep=0pt, outer sep=0pt, scale=  0.88] at (134.37,267.63) {K-Cycles};
\end{scope}
\begin{scope}
\path[clip] (  0.00,  0.00) rectangle (289.08,289.08);
\definecolor{fillColor}{RGB}{255,255,255}

\path[fill=fillColor] (185.95,257.74) rectangle (266.33,283.58);
\end{scope}
\begin{scope}
\path[clip] (  0.00,  0.00) rectangle (289.08,289.08);
\definecolor{drawColor}{RGB}{255,255,255}
\definecolor{fillColor}{gray}{0.95}

\path[draw=drawColor,line width= 0.6pt,line join=round,line cap=round,fill=fillColor] (195.25,263.44) rectangle (209.71,277.89);
\end{scope}
\begin{scope}
\path[clip] (  0.00,  0.00) rectangle (289.08,289.08);
\definecolor{drawColor}{RGB}{0,0,0}

\path[draw=drawColor,line width= 0.6pt,line join=round] (196.70,270.66) -- (208.26,270.66);
\end{scope}
\begin{scope}
\path[clip] (  0.00,  0.00) rectangle (289.08,289.08);
\definecolor{drawColor}{RGB}{0,0,0}

\node[text=drawColor,anchor=base west,inner sep=0pt, outer sep=0pt, scale=  0.88] at (211.51,267.63) {Ideal Scaling};
\end{scope}
\end{tikzpicture}

%% file: efficiency.tex
\begin{tikzpicture}[x=1pt,y=1pt]
\definecolor{fillColor}{RGB}{255,255,255}
\path[use as bounding box,fill=fillColor,fill opacity=0.00] (0,0) rectangle (289.08,289.08);
\begin{scope}
\path[clip] (  0.00,  0.00) rectangle (289.08,289.08);
\definecolor{drawColor}{RGB}{255,255,255}
\definecolor{fillColor}{RGB}{255,255,255}

\path[draw=drawColor,line width= 0.6pt,line join=round,line cap=round,fill=fillColor] (  0.00,  0.00) rectangle (289.08,289.08);
\end{scope}
\begin{scope}
\path[clip] ( 49.42, 32.09) rectangle (283.58,231.91);
\definecolor{fillColor}{gray}{0.92}

\path[fill=fillColor] ( 49.42, 32.09) rectangle (283.58,231.91);
\definecolor{drawColor}{RGB}{255,255,255}

\path[draw=drawColor,line width= 0.3pt,line join=round] ( 49.42, 70.87) --
	(283.58, 70.87);

\path[draw=drawColor,line width= 0.3pt,line join=round] ( 49.42,130.27) --
	(283.58,130.27);

\path[draw=drawColor,line width= 0.3pt,line join=round] ( 49.42,189.67) --
	(283.58,189.67);

\path[draw=drawColor,line width= 0.3pt,line join=round] ( 50.91, 32.09) --
	( 50.91,231.91);

\path[draw=drawColor,line width= 0.3pt,line join=round] ( 55.74, 32.09) --
	( 55.74,231.91);

\path[draw=drawColor,line width= 0.3pt,line join=round] ( 88.50, 32.09) --
	( 88.50,231.91);

\path[draw=drawColor,line width= 0.3pt,line join=round] (105.13, 32.09) --
	(105.13,231.91);

\path[draw=drawColor,line width= 0.3pt,line join=round] (116.94, 32.09) --
	(116.94,231.91);

\path[draw=drawColor,line width= 0.3pt,line join=round] (126.09, 32.09) --
	(126.09,231.91);

\path[draw=drawColor,line width= 0.3pt,line join=round] (133.57, 32.09) --
	(133.57,231.91);

\path[draw=drawColor,line width= 0.3pt,line join=round] (139.90, 32.09) --
	(139.90,231.91);

\path[draw=drawColor,line width= 0.3pt,line join=round] (145.37, 32.09) --
	(145.37,231.91);

\path[draw=drawColor,line width= 0.3pt,line join=round] (150.21, 32.09) --
	(150.21,231.91);

\path[draw=drawColor,line width= 0.3pt,line join=round] (182.96, 32.09) --
	(182.96,231.91);

\path[draw=drawColor,line width= 0.3pt,line join=round] (199.60, 32.09) --
	(199.60,231.91);

\path[draw=drawColor,line width= 0.3pt,line join=round] (211.40, 32.09) --
	(211.40,231.91);

\path[draw=drawColor,line width= 0.3pt,line join=round] (220.56, 32.09) --
	(220.56,231.91);

\path[draw=drawColor,line width= 0.3pt,line join=round] (228.04, 32.09) --
	(228.04,231.91);

\path[draw=drawColor,line width= 0.3pt,line join=round] (234.36, 32.09) --
	(234.36,231.91);

\path[draw=drawColor,line width= 0.3pt,line join=round] (239.84, 32.09) --
	(239.84,231.91);

\path[draw=drawColor,line width= 0.3pt,line join=round] (244.67, 32.09) --
	(244.67,231.91);

\path[draw=drawColor,line width= 0.3pt,line join=round] (277.43, 32.09) --
	(277.43,231.91);

\path[draw=drawColor,line width= 0.6pt,line join=round] ( 49.42, 41.17) --
	(283.58, 41.17);

\path[draw=drawColor,line width= 0.6pt,line join=round] ( 49.42,100.57) --
	(283.58,100.57);

\path[draw=drawColor,line width= 0.6pt,line join=round] ( 49.42,159.97) --
	(283.58,159.97);

\path[draw=drawColor,line width= 0.6pt,line join=round] ( 49.42,219.36) --
	(283.58,219.36);

\path[draw=drawColor,line width= 0.6pt,line join=round] ( 60.06, 32.09) --
	( 60.06,231.91);

\path[draw=drawColor,line width= 0.6pt,line join=round] (154.53, 32.09) --
	(154.53,231.91);

\path[draw=drawColor,line width= 0.6pt,line join=round] (248.99, 32.09) --
	(248.99,231.91);
\definecolor{fillColor}{RGB}{0,0,0}

\path[fill=fillColor] (272.94, 72.89) circle (  3.57);

\path[fill=fillColor] (259.04, 75.69) circle (  3.57);

\path[fill=fillColor] (250.40, 78.12) circle (  3.57);

\path[fill=fillColor] (245.77, 74.57) circle (  3.57);

\path[fill=fillColor] (232.09,178.77) --
	(236.90,170.44) --
	(227.28,170.44) --
	cycle;

\path[fill=fillColor] (189.61,141.04) --
	(194.42,132.72) --
	(184.80,132.72) --
	cycle;

\path[fill=fillColor] (164.75,123.19) --
	(169.55,114.86) --
	(159.94,114.86) --
	cycle;

\path[fill=fillColor] (139.97,126.01) --
	(144.78,117.68) --
	(135.17,117.68) --
	cycle;

\path[fill=fillColor] (131.89,108.04) --
	(136.70, 99.72) --
	(127.09, 99.72) --
	cycle;

\path[fill=fillColor] (119.07,219.26) --
	(126.20,219.26) --
	(126.20,226.40) --
	(119.07,226.40) --
	cycle;

\path[fill=fillColor] (106.71,208.60) --
	(113.84,208.60) --
	(113.84,215.74) --
	(106.71,215.74) --
	cycle;

\path[fill=fillColor] (103.68,193.58) --
	(110.82,193.58) --
	(110.82,200.71) --
	(103.68,200.71) --
	cycle;

\path[fill=fillColor] ( 90.16,199.13) --
	( 97.30,199.13) --
	( 97.30,206.27) --
	( 90.16,206.27) --
	cycle;
\definecolor{drawColor}{RGB}{0,0,0}

\path[draw=drawColor,line width= 0.4pt,line join=round,line cap=round] (132.23,171.14) -- (142.32,171.14);

\path[draw=drawColor,line width= 0.4pt,line join=round,line cap=round] (137.27,166.09) -- (137.27,176.18);

\path[draw=drawColor,line width= 0.4pt,line join=round,line cap=round] (108.88,187.74) -- (118.97,187.74);

\path[draw=drawColor,line width= 0.4pt,line join=round,line cap=round] (113.92,182.69) -- (113.92,192.79);

\path[draw=drawColor,line width= 0.4pt,line join=round,line cap=round] (114.98,129.17) -- (125.08,129.17);

\path[draw=drawColor,line width= 0.4pt,line join=round,line cap=round] (120.03,124.12) -- (120.03,134.22);

\path[draw=drawColor,line width= 0.4pt,line join=round,line cap=round] ( 92.12,155.81) -- (102.21,155.81);

\path[draw=drawColor,line width= 0.4pt,line join=round,line cap=round] ( 97.17,150.76) -- ( 97.17,160.86);

\path[draw=drawColor,line width= 0.6pt,line join=round,line cap=round] (272.25, 55.70) -- (272.86, 71.08);

\node[text=drawColor,anchor=base,inner sep=0pt, outer sep=0pt, scale=  1.14] at (272.04, 46.36) {324};

\node[text=drawColor,anchor=base,inner sep=0pt, outer sep=0pt, scale=  1.14] at (257.50, 60.18) {400};

\node[text=drawColor,anchor=base,inner sep=0pt, outer sep=0pt, scale=  1.14] at (250.90, 84.80) {484};

\node[text=drawColor,anchor=base,inner sep=0pt, outer sep=0pt, scale=  1.14] at (232.98, 60.04) {576};

\node[text=drawColor,anchor=base,inner sep=0pt, outer sep=0pt, scale=  1.14] at (241.56,177.38) {4};

\node[text=drawColor,anchor=base,inner sep=0pt, outer sep=0pt, scale=  1.14] at (201.96,142.12) {16};

\node[text=drawColor,anchor=base,inner sep=0pt, outer sep=0pt, scale=  1.14] at (177.09,124.25) {36};

\node[text=drawColor,anchor=base,inner sep=0pt, outer sep=0pt, scale=  1.14] at (152.29,106.06) {64};

\node[text=drawColor,anchor=base,inner sep=0pt, outer sep=0pt, scale=  1.14] at (119.51, 88.01) {100};

\node[text=drawColor,anchor=base,inner sep=0pt, outer sep=0pt, scale=  1.14] at (137.80,221.06) {144};

\node[text=drawColor,anchor=base,inner sep=0pt, outer sep=0pt, scale=  1.14] at (101.34,205.34) {196};

\path[draw=drawColor,line width= 0.6pt,line join=round,line cap=round] (100.33,191.98) -- (105.45,195.80);

\node[text=drawColor,anchor=base,inner sep=0pt, outer sep=0pt, scale=  1.14] at ( 93.06,182.64) {256};

\path[draw=drawColor,line width= 0.6pt,line join=round,line cap=round] ( 85.29,208.92) -- ( 91.93,204.03);

\node[text=drawColor,anchor=base,inner sep=0pt, outer sep=0pt, scale=  1.14] at ( 77.94,210.43) {324};

\node[text=drawColor,anchor=base,inner sep=0pt, outer sep=0pt, scale=  1.14] at (149.62,156.68) {64};

\node[text=drawColor,anchor=base,inner sep=0pt, outer sep=0pt, scale=  1.14] at (126.24,176.59) {100};

\node[text=drawColor,anchor=base,inner sep=0pt, outer sep=0pt, scale=  1.14] at (107.66,135.81) {144};

\node[text=drawColor,anchor=base,inner sep=0pt, outer sep=0pt, scale=  1.14] at (109.53,162.44) {196};
\end{scope}
\begin{scope}
\path[clip] (  0.00,  0.00) rectangle (289.08,289.08);
\definecolor{drawColor}{gray}{0.30}

\node[text=drawColor,anchor=base east,inner sep=0pt, outer sep=0pt, scale=  0.88] at ( 44.47, 38.14) {0e+00};

\node[text=drawColor,anchor=base east,inner sep=0pt, outer sep=0pt, scale=  0.88] at ( 44.47, 97.54) {1e+06};

\node[text=drawColor,anchor=base east,inner sep=0pt, outer sep=0pt, scale=  0.88] at ( 44.47,156.94) {2e+06};

\node[text=drawColor,anchor=base east,inner sep=0pt, outer sep=0pt, scale=  0.88] at ( 44.47,216.33) {3e+06};
\end{scope}
\begin{scope}
\path[clip] (  0.00,  0.00) rectangle (289.08,289.08);
\definecolor{drawColor}{gray}{0.20}

\path[draw=drawColor,line width= 0.6pt,line join=round] ( 46.67, 41.17) --
	( 49.42, 41.17);

\path[draw=drawColor,line width= 0.6pt,line join=round] ( 46.67,100.57) --
	( 49.42,100.57);

\path[draw=drawColor,line width= 0.6pt,line join=round] ( 46.67,159.97) --
	( 49.42,159.97);

\path[draw=drawColor,line width= 0.6pt,line join=round] ( 46.67,219.36) --
	( 49.42,219.36);
\end{scope}
\begin{scope}
\path[clip] (  0.00,  0.00) rectangle (289.08,289.08);
\definecolor{drawColor}{gray}{0.20}

\path[draw=drawColor,line width= 0.6pt,line join=round] ( 60.06, 29.34) --
	( 60.06, 32.09);

\path[draw=drawColor,line width= 0.6pt,line join=round] (154.53, 29.34) --
	(154.53, 32.09);

\path[draw=drawColor,line width= 0.6pt,line join=round] (248.99, 29.34) --
	(248.99, 32.09);
\end{scope}
\begin{scope}
\path[clip] (  0.00,  0.00) rectangle (289.08,289.08);
\definecolor{drawColor}{gray}{0.30}

\node[text=drawColor,anchor=base,inner sep=0pt, outer sep=0pt, scale=  0.88] at ( 60.06, 21.08) {1};

\node[text=drawColor,anchor=base,inner sep=0pt, outer sep=0pt, scale=  0.88] at (154.53, 21.08) {10};

\node[text=drawColor,anchor=base,inner sep=0pt, outer sep=0pt, scale=  0.88] at (248.99, 21.08) {100};
\end{scope}
\begin{scope}
\path[clip] (  0.00,  0.00) rectangle (289.08,289.08);
\definecolor{drawColor}{RGB}{0,0,0}

\node[text=drawColor,anchor=base,inner sep=0pt, outer sep=0pt, scale=  1.10] at (166.50,  8.00) {Solve Time (seconds)};
\end{scope}
\begin{scope}
\path[clip] (  0.00,  0.00) rectangle (289.08,289.08);
\definecolor{drawColor}{RGB}{0,0,0}

\node[text=drawColor,rotate= 90.00,anchor=base,inner sep=0pt, outer sep=0pt, scale=  1.10] at ( 13.08,132.00) {Efficiency per Digit of Accuracy};
\end{scope}
\begin{scope}
\path[clip] (  0.00,  0.00) rectangle (289.08,289.08);
\definecolor{fillColor}{RGB}{255,255,255}

\path[fill=fillColor] ( 29.89,243.29) rectangle (303.11,283.58);
\end{scope}
\begin{scope}
\path[clip] (  0.00,  0.00) rectangle (289.08,289.08);
\definecolor{drawColor}{RGB}{0,0,0}

\node[text=drawColor,anchor=base west,inner sep=0pt, outer sep=0pt, scale=  1.10] at ( 35.58,259.65) {Graph (NNZ)};
\end{scope}
\begin{scope}
\path[clip] (  0.00,  0.00) rectangle (289.08,289.08);
\definecolor{drawColor}{RGB}{255,255,255}
\definecolor{fillColor}{gray}{0.95}

\path[draw=drawColor,line width= 0.6pt,line join=round,line cap=round,fill=fillColor] (105.28,263.44) rectangle (119.74,277.89);
\end{scope}
\begin{scope}
\path[clip] (  0.00,  0.00) rectangle (289.08,289.08);
\definecolor{fillColor}{RGB}{0,0,0}

\path[fill=fillColor] (112.51,270.66) circle (  3.57);
\end{scope}
\begin{scope}
\path[clip] (  0.00,  0.00) rectangle (289.08,289.08);
\definecolor{drawColor}{RGB}{255,255,255}
\definecolor{fillColor}{gray}{0.95}

\path[draw=drawColor,line width= 0.6pt,line join=round,line cap=round,fill=fillColor] (105.28,248.98) rectangle (119.74,263.44);
\end{scope}
\begin{scope}
\path[clip] (  0.00,  0.00) rectangle (289.08,289.08);
\definecolor{fillColor}{RGB}{0,0,0}

\path[fill=fillColor] (112.51,261.76) --
	(117.32,253.43) --
	(107.70,253.43) --
	cycle;
\end{scope}
\begin{scope}
\path[clip] (  0.00,  0.00) rectangle (289.08,289.08);
\definecolor{drawColor}{RGB}{255,255,255}
\definecolor{fillColor}{gray}{0.95}

\path[draw=drawColor,line width= 0.6pt,line join=round,line cap=round,fill=fillColor] (210.75,263.44) rectangle (225.21,277.89);
\end{scope}
\begin{scope}
\path[clip] (  0.00,  0.00) rectangle (289.08,289.08);
\definecolor{fillColor}{RGB}{0,0,0}

\path[fill=fillColor] (214.41,267.09) --
	(221.55,267.09) --
	(221.55,274.23) --
	(214.41,274.23) --
	cycle;
\end{scope}
\begin{scope}
\path[clip] (  0.00,  0.00) rectangle (289.08,289.08);
\definecolor{drawColor}{RGB}{255,255,255}
\definecolor{fillColor}{gray}{0.95}

\path[draw=drawColor,line width= 0.6pt,line join=round,line cap=round,fill=fillColor] (210.75,248.98) rectangle (225.21,263.44);
\end{scope}
\begin{scope}
\path[clip] (  0.00,  0.00) rectangle (289.08,289.08);
\definecolor{drawColor}{RGB}{0,0,0}

\path[draw=drawColor,line width= 0.4pt,line join=round,line cap=round] (212.93,256.21) -- (223.03,256.21);

\path[draw=drawColor,line width= 0.4pt,line join=round,line cap=round] (217.98,251.16) -- (217.98,261.26);
\end{scope}
\begin{scope}
\path[clip] (  0.00,  0.00) rectangle (289.08,289.08);
\definecolor{drawColor}{RGB}{0,0,0}

\node[text=drawColor,anchor=base west,inner sep=0pt, outer sep=0pt, scale=  0.88] at (121.54,267.63) {com-friendster (3.61G)};
\end{scope}
\begin{scope}
\path[clip] (  0.00,  0.00) rectangle (289.08,289.08);
\definecolor{drawColor}{RGB}{0,0,0}

\node[text=drawColor,anchor=base west,inner sep=0pt, outer sep=0pt, scale=  0.88] at (121.54,253.18) {com-lj (69.4M)};
\end{scope}
\begin{scope}
\path[clip] (  0.00,  0.00) rectangle (289.08,289.08);
\definecolor{drawColor}{RGB}{0,0,0}

\node[text=drawColor,anchor=base west,inner sep=0pt, outer sep=0pt, scale=  0.88] at (227.01,267.63) {com-orkut (234M)};
\end{scope}
\begin{scope}
\path[clip] (  0.00,  0.00) rectangle (289.08,289.08);
\definecolor{drawColor}{RGB}{0,0,0}

\node[text=drawColor,anchor=base west,inner sep=0pt, outer sep=0pt, scale=  0.88] at (227.01,253.18) {hollywood (114M)};
\end{scope}
\end{tikzpicture}

%% file: efficiency_wda.tex
\begin{tikzpicture}[x=1pt,y=1pt]
\definecolor{fillColor}{RGB}{255,255,255}
\path[use as bounding box,fill=fillColor,fill opacity=0.00] (0,0) rectangle (289.08,289.08);
\begin{scope}
\path[clip] (  0.00,  0.00) rectangle (289.08,289.08);
\definecolor{drawColor}{RGB}{255,255,255}
\definecolor{fillColor}{RGB}{255,255,255}

\path[draw=drawColor,line width= 0.6pt,line join=round,line cap=round,fill=fillColor] (  0.00,  0.00) rectangle (289.08,289.08);
\end{scope}
\begin{scope}
\path[clip] ( 56.26, 31.53) rectangle (283.58,231.91);
\definecolor{fillColor}{gray}{0.92}

\path[fill=fillColor] ( 56.26, 31.53) rectangle (283.58,231.91);
\definecolor{drawColor}{RGB}{255,255,255}

\path[draw=drawColor,line width= 0.3pt,line join=round] ( 56.26, 65.98) --
	(283.58, 65.98);

\path[draw=drawColor,line width= 0.3pt,line join=round] ( 56.26,116.67) --
	(283.58,116.67);

\path[draw=drawColor,line width= 0.3pt,line join=round] ( 56.26,167.37) --
	(283.58,167.37);

\path[draw=drawColor,line width= 0.3pt,line join=round] ( 56.26,218.06) --
	(283.58,218.06);

\path[draw=drawColor,line width= 0.3pt,line join=round] ( 58.98, 31.53) --
	( 58.98,231.91);

\path[draw=drawColor,line width= 0.3pt,line join=round] ( 66.60, 31.53) --
	( 66.60,231.91);

\path[draw=drawColor,line width= 0.3pt,line join=round] ( 72.82, 31.53) --
	( 72.82,231.91);

\path[draw=drawColor,line width= 0.3pt,line join=round] ( 78.08, 31.53) --
	( 78.08,231.91);

\path[draw=drawColor,line width= 0.3pt,line join=round] ( 82.64, 31.53) --
	( 82.64,231.91);

\path[draw=drawColor,line width= 0.3pt,line join=round] ( 86.66, 31.53) --
	( 86.66,231.91);

\path[draw=drawColor,line width= 0.3pt,line join=round] (113.93, 31.53) --
	(113.93,231.91);

\path[draw=drawColor,line width= 0.3pt,line join=round] (127.77, 31.53) --
	(127.77,231.91);

\path[draw=drawColor,line width= 0.3pt,line join=round] (137.60, 31.53) --
	(137.60,231.91);

\path[draw=drawColor,line width= 0.3pt,line join=round] (145.21, 31.53) --
	(145.21,231.91);

\path[draw=drawColor,line width= 0.3pt,line join=round] (151.44, 31.53) --
	(151.44,231.91);

\path[draw=drawColor,line width= 0.3pt,line join=round] (156.70, 31.53) --
	(156.70,231.91);

\path[draw=drawColor,line width= 0.3pt,line join=round] (161.26, 31.53) --
	(161.26,231.91);

\path[draw=drawColor,line width= 0.3pt,line join=round] (165.28, 31.53) --
	(165.28,231.91);

\path[draw=drawColor,line width= 0.3pt,line join=round] (192.55, 31.53) --
	(192.55,231.91);

\path[draw=drawColor,line width= 0.3pt,line join=round] (206.39, 31.53) --
	(206.39,231.91);

\path[draw=drawColor,line width= 0.3pt,line join=round] (216.21, 31.53) --
	(216.21,231.91);

\path[draw=drawColor,line width= 0.3pt,line join=round] (223.83, 31.53) --
	(223.83,231.91);

\path[draw=drawColor,line width= 0.3pt,line join=round] (230.06, 31.53) --
	(230.06,231.91);

\path[draw=drawColor,line width= 0.3pt,line join=round] (235.32, 31.53) --
	(235.32,231.91);

\path[draw=drawColor,line width= 0.3pt,line join=round] (239.88, 31.53) --
	(239.88,231.91);

\path[draw=drawColor,line width= 0.3pt,line join=round] (243.90, 31.53) --
	(243.90,231.91);

\path[draw=drawColor,line width= 0.3pt,line join=round] (271.17, 31.53) --
	(271.17,231.91);

\path[draw=drawColor,line width= 0.6pt,line join=round] ( 56.26, 40.64) --
	(283.58, 40.64);

\path[draw=drawColor,line width= 0.6pt,line join=round] ( 56.26, 91.33) --
	(283.58, 91.33);

\path[draw=drawColor,line width= 0.6pt,line join=round] ( 56.26,142.02) --
	(283.58,142.02);

\path[draw=drawColor,line width= 0.6pt,line join=round] ( 56.26,192.71) --
	(283.58,192.71);

\path[draw=drawColor,line width= 0.6pt,line join=round] ( 90.26, 31.53) --
	( 90.26,231.91);

\path[draw=drawColor,line width= 0.6pt,line join=round] (168.88, 31.53) --
	(168.88,231.91);

\path[draw=drawColor,line width= 0.6pt,line join=round] (247.50, 31.53) --
	(247.50,231.91);
\definecolor{drawColor}{RGB}{0,0,0}

\path[draw=drawColor,line width= 0.6pt,line join=round,line cap=round] (267.64, 71.63) -- (271.44, 75.07);

\node[text=drawColor,anchor=base,inner sep=0pt, outer sep=0pt, scale=  1.14] at (261.64, 62.29) {324};

\path[draw=drawColor,line width= 0.6pt,line join=round,line cap=round] (267.42, 83.98) -- (264.68, 80.74);

\node[text=drawColor,anchor=base,inner sep=0pt, outer sep=0pt, scale=  1.14] at (271.99, 85.48) {400};

\path[draw=drawColor,line width= 0.6pt,line join=round,line cap=round] (248.52, 74.27) -- (252.53, 76.28);

\node[text=drawColor,anchor=base,inner sep=0pt, outer sep=0pt, scale=  1.14] at (238.48, 65.31) {484};

\path[draw=drawColor,line width= 0.6pt,line join=round,line cap=round] (245.99, 85.03) -- (250.03, 81.47);

\node[text=drawColor,anchor=base,inner sep=0pt, outer sep=0pt, scale=  1.14] at (239.82, 86.53) {576};

\path[draw=drawColor,line width= 0.6pt,line join=round,line cap=round] (235.76,218.29) -- (239.12,221.22);

\node[text=drawColor,anchor=base,inner sep=0pt, outer sep=0pt, scale=  1.14] at (231.41,210.57) {4};

\node[text=drawColor,anchor=base,inner sep=0pt, outer sep=0pt, scale=  1.14] at (199.90,154.12) {16};

\node[text=drawColor,anchor=base,inner sep=0pt, outer sep=0pt, scale=  1.14] at (189.83,149.22) {36};

\path[draw=drawColor,line width= 0.6pt,line join=round,line cap=round] (170.82,130.50) -- (165.49,132.24);

\node[text=drawColor,anchor=base,inner sep=0pt, outer sep=0pt, scale=  1.14] at (178.02,124.23) {64};

\path[draw=drawColor,line width= 0.6pt,line join=round,line cap=round] (163.17, 93.99) -- (159.02, 97.68);

\node[text=drawColor,anchor=base,inner sep=0pt, outer sep=0pt, scale=  1.14] at (169.25, 84.64) {100};

\path[draw=drawColor,line width= 0.6pt,line join=round,line cap=round] (154.54,125.13) -- (148.86,134.73);

\node[text=drawColor,anchor=base,inner sep=0pt, outer sep=0pt, scale=  1.14] at (157.75,115.79) {144};

\node[text=drawColor,anchor=base,inner sep=0pt, outer sep=0pt, scale=  1.14] at (132.74,127.60) {196};

\path[draw=drawColor,line width= 0.6pt,line join=round,line cap=round] (134.47,111.15) -- (133.60,117.37);

\node[text=drawColor,anchor=base,inner sep=0pt, outer sep=0pt, scale=  1.14] at (135.23,101.81) {256};

\node[text=drawColor,anchor=base,inner sep=0pt, outer sep=0pt, scale=  1.14] at (108.71,114.25) {324};

\path[draw=drawColor,line width= 0.6pt,line join=round,line cap=round] (167.62,186.90) -- (163.57,183.31);

\node[text=drawColor,anchor=base,inner sep=0pt, outer sep=0pt, scale=  1.14] at (173.73,188.40) {64};

\path[draw=drawColor,line width= 0.6pt,line join=round,line cap=round] (136.41,174.64) -- (140.61,178.39);

\node[text=drawColor,anchor=base,inner sep=0pt, outer sep=0pt, scale=  1.14] at (130.32,165.29) {100};

\node[text=drawColor,anchor=base,inner sep=0pt, outer sep=0pt, scale=  1.14] at (146.45,146.77) {144};

\path[draw=drawColor,line width= 0.6pt,line join=round,line cap=round] (121.87,142.91) -- (126.03,139.25);

\node[text=drawColor,anchor=base,inner sep=0pt, outer sep=0pt, scale=  1.14] at (115.73,144.42) {196};
\definecolor{fillColor}{RGB}{0,0,0}

\path[fill=fillColor] (273.25, 76.70) circle (  3.57);

\path[fill=fillColor] (263.16, 78.93) circle (  3.57);

\path[fill=fillColor] (254.34, 77.18) circle (  3.57);

\path[fill=fillColor] (251.83, 79.88) circle (  3.57);

\path[fill=fillColor] (240.93,228.35) --
	(245.74,220.03) --
	(236.13,220.03) --
	cycle;

\path[fill=fillColor] (205.08,168.14) --
	(209.89,159.82) --
	(200.28,159.82) --
	cycle;

\path[fill=fillColor] (184.56,154.29) --
	(189.37,145.97) --
	(179.76,145.97) --
	cycle;

\path[fill=fillColor] (163.68,138.38) --
	(168.49,130.05) --
	(158.87,130.05) --
	cycle;

\path[fill=fillColor] (157.21,104.84) --
	(162.02, 96.52) --
	(152.41, 96.52) --
	cycle;

\path[fill=fillColor] (144.22,132.97) --
	(151.36,132.97) --
	(151.36,140.11) --
	(144.22,140.11) --
	cycle;

\path[fill=fillColor] (135.76,122.97) --
	(142.90,122.97) --
	(142.90,130.10) --
	(135.76,130.10) --
	cycle;

\path[fill=fillColor] (129.78,115.61) --
	(136.92,115.61) --
	(136.92,122.75) --
	(129.78,122.75) --
	cycle;

\path[fill=fillColor] (120.54,114.86) --
	(127.68,114.86) --
	(127.68,122.00) --
	(120.54,122.00) --
	cycle;

\path[draw=drawColor,line width= 0.4pt,line join=round,line cap=round] (156.72,181.70) -- (166.81,181.70);

\path[draw=drawColor,line width= 0.4pt,line join=round,line cap=round] (161.76,176.66) -- (161.76,186.75);

\path[draw=drawColor,line width= 0.4pt,line join=round,line cap=round] (137.37,180.00) -- (147.47,180.00);

\path[draw=drawColor,line width= 0.4pt,line join=round,line cap=round] (142.42,174.95) -- (142.42,185.05);

\path[draw=drawColor,line width= 0.4pt,line join=round,line cap=round] (142.34,140.08) -- (152.43,140.08);

\path[draw=drawColor,line width= 0.4pt,line join=round,line cap=round] (147.39,135.04) -- (147.39,145.13);

\path[draw=drawColor,line width= 0.4pt,line join=round,line cap=round] (122.79,137.65) -- (132.88,137.65);

\path[draw=drawColor,line width= 0.4pt,line join=round,line cap=round] (127.83,132.60) -- (127.83,142.70);
\end{scope}
\begin{scope}
\path[clip] (  0.00,  0.00) rectangle (289.08,289.08);
\definecolor{drawColor}{gray}{0.30}

\node[text=drawColor,anchor=base east,inner sep=0pt, outer sep=0pt, scale=  0.88] at ( 51.31, 37.61) {0.0e+00};

\node[text=drawColor,anchor=base east,inner sep=0pt, outer sep=0pt, scale=  0.88] at ( 51.31, 88.30) {2.5e+07};

\node[text=drawColor,anchor=base east,inner sep=0pt, outer sep=0pt, scale=  0.88] at ( 51.31,138.99) {5.0e+07};

\node[text=drawColor,anchor=base east,inner sep=0pt, outer sep=0pt, scale=  0.88] at ( 51.31,189.68) {7.5e+07};
\end{scope}
\begin{scope}
\path[clip] (  0.00,  0.00) rectangle (289.08,289.08);
\definecolor{drawColor}{gray}{0.20}

\path[draw=drawColor,line width= 0.6pt,line join=round] ( 53.51, 40.64) --
	( 56.26, 40.64);

\path[draw=drawColor,line width= 0.6pt,line join=round] ( 53.51, 91.33) --
	( 56.26, 91.33);

\path[draw=drawColor,line width= 0.6pt,line join=round] ( 53.51,142.02) --
	( 56.26,142.02);

\path[draw=drawColor,line width= 0.6pt,line join=round] ( 53.51,192.71) --
	( 56.26,192.71);
\end{scope}
\begin{scope}
\path[clip] (  0.00,  0.00) rectangle (289.08,289.08);
\definecolor{drawColor}{gray}{0.20}

\path[draw=drawColor,line width= 0.6pt,line join=round] ( 90.26, 28.78) --
	( 90.26, 31.53);

\path[draw=drawColor,line width= 0.6pt,line join=round] (168.88, 28.78) --
	(168.88, 31.53);

\path[draw=drawColor,line width= 0.6pt,line join=round] (247.50, 28.78) --
	(247.50, 31.53);
\end{scope}
\begin{scope}
\path[clip] (  0.00,  0.00) rectangle (289.08,289.08);
\definecolor{drawColor}{gray}{0.30}

\node[text=drawColor,anchor=base,inner sep=0pt, outer sep=0pt, scale=  0.88] at ( 90.26, 20.52) {0.1};

\node[text=drawColor,anchor=base,inner sep=0pt, outer sep=0pt, scale=  0.88] at (168.88, 20.52) {1.0};

\node[text=drawColor,anchor=base,inner sep=0pt, outer sep=0pt, scale=  0.88] at (247.50, 20.52) {10.0};
\end{scope}
\begin{scope}
\path[clip] (  0.00,  0.00) rectangle (289.08,289.08);
\definecolor{drawColor}{RGB}{0,0,0}

\node[text=drawColor,anchor=base,inner sep=0pt, outer sep=0pt, scale=  1.10] at (169.92,  7.44) {Time per Digit of Accuracy};
\end{scope}
\begin{scope}
\path[clip] (  0.00,  0.00) rectangle (289.08,289.08);
\definecolor{drawColor}{RGB}{0,0,0}

\node[text=drawColor,rotate= 90.00,anchor=base,inner sep=0pt, outer sep=0pt, scale=  1.10] at ( 13.08,131.72) {Normalized Efficiency};
\end{scope}
\begin{scope}
\path[clip] (  0.00,  0.00) rectangle (289.08,289.08);
\definecolor{fillColor}{RGB}{255,255,255}

\path[fill=fillColor] ( 33.31,243.29) rectangle (306.53,283.58);
\end{scope}
\begin{scope}
\path[clip] (  0.00,  0.00) rectangle (289.08,289.08);
\definecolor{drawColor}{RGB}{0,0,0}

\node[text=drawColor,anchor=base west,inner sep=0pt, outer sep=0pt, scale=  1.10] at ( 39.00,259.65) {Graph (NNZ)};
\end{scope}
\begin{scope}
\path[clip] (  0.00,  0.00) rectangle (289.08,289.08);
\definecolor{drawColor}{RGB}{255,255,255}
\definecolor{fillColor}{gray}{0.95}

\path[draw=drawColor,line width= 0.6pt,line join=round,line cap=round,fill=fillColor] (108.70,263.44) rectangle (123.16,277.89);
\end{scope}
\begin{scope}
\path[clip] (  0.00,  0.00) rectangle (289.08,289.08);
\definecolor{fillColor}{RGB}{0,0,0}

\path[fill=fillColor] (115.93,270.66) circle (  3.57);
\end{scope}
\begin{scope}
\path[clip] (  0.00,  0.00) rectangle (289.08,289.08);
\definecolor{drawColor}{RGB}{255,255,255}
\definecolor{fillColor}{gray}{0.95}

\path[draw=drawColor,line width= 0.6pt,line join=round,line cap=round,fill=fillColor] (108.70,248.98) rectangle (123.16,263.44);
\end{scope}
\begin{scope}
\path[clip] (  0.00,  0.00) rectangle (289.08,289.08);
\definecolor{fillColor}{RGB}{0,0,0}

\path[fill=fillColor] (115.93,261.76) --
	(120.74,253.43) --
	(111.12,253.43) --
	cycle;
\end{scope}
\begin{scope}
\path[clip] (  0.00,  0.00) rectangle (289.08,289.08);
\definecolor{drawColor}{RGB}{255,255,255}
\definecolor{fillColor}{gray}{0.95}

\path[draw=drawColor,line width= 0.6pt,line join=round,line cap=round,fill=fillColor] (214.18,263.44) rectangle (228.63,277.89);
\end{scope}
\begin{scope}
\path[clip] (  0.00,  0.00) rectangle (289.08,289.08);
\definecolor{fillColor}{RGB}{0,0,0}

\path[fill=fillColor] (217.83,267.09) --
	(224.97,267.09) --
	(224.97,274.23) --
	(217.83,274.23) --
	cycle;
\end{scope}
\begin{scope}
\path[clip] (  0.00,  0.00) rectangle (289.08,289.08);
\definecolor{drawColor}{RGB}{255,255,255}
\definecolor{fillColor}{gray}{0.95}

\path[draw=drawColor,line width= 0.6pt,line join=round,line cap=round,fill=fillColor] (214.18,248.98) rectangle (228.63,263.44);
\end{scope}
\begin{scope}
\path[clip] (  0.00,  0.00) rectangle (289.08,289.08);
\definecolor{drawColor}{RGB}{0,0,0}

\path[draw=drawColor,line width= 0.4pt,line join=round,line cap=round] (216.36,256.21) -- (226.45,256.21);

\path[draw=drawColor,line width= 0.4pt,line join=round,line cap=round] (221.40,251.16) -- (221.40,261.26);
\end{scope}
\begin{scope}
\path[clip] (  0.00,  0.00) rectangle (289.08,289.08);
\definecolor{drawColor}{RGB}{0,0,0}

\node[text=drawColor,anchor=base west,inner sep=0pt, outer sep=0pt, scale=  0.88] at (124.96,267.63) {com-friendster (3.61G)};
\end{scope}
\begin{scope}
\path[clip] (  0.00,  0.00) rectangle (289.08,289.08);
\definecolor{drawColor}{RGB}{0,0,0}

\node[text=drawColor,anchor=base west,inner sep=0pt, outer sep=0pt, scale=  0.88] at (124.96,253.18) {com-lj (69.4M)};
\end{scope}
\begin{scope}
\path[clip] (  0.00,  0.00) rectangle (289.08,289.08);
\definecolor{drawColor}{RGB}{0,0,0}

\node[text=drawColor,anchor=base west,inner sep=0pt, outer sep=0pt, scale=  0.88] at (230.44,267.63) {com-orkut (234M)};
\end{scope}
\begin{scope}
\path[clip] (  0.00,  0.00) rectangle (289.08,289.08);
\definecolor{drawColor}{RGB}{0,0,0}

\node[text=drawColor,anchor=base west,inner sep=0pt, outer sep=0pt, scale=  0.88] at (230.44,253.18) {hollywood (114M)};
\end{scope}
\end{tikzpicture}